\documentclass[10pt, conference]{IEEEtran}
\IEEEoverridecommandlockouts
\usepackage{cite}
\usepackage{amsmath,amssymb,amsfonts}
\usepackage{graphicx}
\usepackage{textcomp}
\usepackage{xcolor}
\usepackage{xcolor, soul}
\usepackage{tcolorbox}
\usepackage{colortbl}
\usepackage{diagbox}
\usepackage{bm}

\usepackage{xcolor,colortbl}
\usepackage{float}
\usepackage{graphicx}
\usepackage{subfigure}
\usepackage{multirow}
\usepackage{xcolor}
\usepackage{colortbl}
\usepackage[bookmarks=false]{hyperref}
\usepackage[ruled,vlined]{algorithm2e}
\usepackage{algpseudocode}
\usepackage{hyperref}
\usepackage{framed}
\usepackage{amsmath}

\usepackage{pifont}
\usepackage{enumitem}
\usepackage{balance}
\usepackage{url}
\usepackage[T1]{fontenc}
\usepackage[utf8]{inputenc}
\usepackage{booktabs}
\usepackage{threeparttable}
\usepackage{xcolor}

\definecolor{grey}{rgb}{0.9,0.9,0.9}
\definecolor{lightgreen}{HTML}{bae4b3}
\definecolor{lightgrey}{HTML}{F0F0F0}
\definecolor{mygreen}{HTML}{31a354}
\definecolor{mygray}{HTML}{666666}

\hypersetup{
	colorlinks=true,      
	linkcolor=black,       
	citecolor=black,    
	filecolor=cyan,       
	urlcolor=black          
}

\newcommand{\scc}[1]{\mytodoblue{[scc: #1]}}
\newcommand{\ying}[1]{\mytodored{[ying: #1]}}

\newcommand{\mytodoblue}[1]{\textcolor{blue}{\ding{46}~{\sf}~#1}}
\newcommand{\mytodored}[1]{\textcolor{red}{\ding{46}~{\sf}~#1}}

\newcommand*{\mycode}{\fontfamily{lmtt}\selectfont}

\definecolor{orange}{RGB}{255,182,193} 

\begin{document}

\title{\textsc{Hero}: On the Chaos When PATH Meets Modules}

\author{\IEEEauthorblockN{Ying Wang\IEEEauthorrefmark{1},
		Liang Qiao\IEEEauthorrefmark{1},
		Chang Xu\IEEEauthorrefmark{2}\IEEEauthorrefmark{4}\thanks{\IEEEauthorrefmark{4}\small{Chang Xu is the corresponding author.}},
		Yepang Liu\IEEEauthorrefmark{3},
		Shing-Chi Cheung\IEEEauthorrefmark{5},
		Na Meng\IEEEauthorrefmark{6},\\
		Hai Yu\IEEEauthorrefmark{1}, and
		Zhiliang Zhu\IEEEauthorrefmark{1}}

	\IEEEauthorblockA{\IEEEauthorrefmark{1} Software College, Northeastern University, China
		\\ Email: wangying@swc.neu.edu.cn, qiaoliangneu@163.com, \{yuhai, zzl\}@mail.neu.edu.cn}
		\IEEEauthorblockA{\IEEEauthorrefmark{2}	State Key Laboratory for Novel Software Technology and Department of Computer Science and Technology, \\Nanjing University, China, Email: changxu@nju.edu.cn}
		\IEEEauthorblockA{\IEEEauthorrefmark{3} Southern University of Science and Technology, China, Email: liuyp1@sustech.edu.cn}
		\IEEEauthorblockA{\IEEEauthorrefmark{5}	The Hong Kong University of Science and Technology, China, Email: scc@cse.ust.hk}
		\IEEEauthorblockA{\IEEEauthorrefmark{6} Virginia Tech, USA, Email: nm8247@cs.vt.edu}

}
\vspace{-8mm}
\maketitle
\newcommand{\codefont}[1]{\footnotesize{\texttt{#1}}\normalsize}
\newcommand{\tool}{\textsc{Hero}\xspace}
\newcommand{\todo}[2]{{\textcolor{#1} {\textbf{#2}}}}

\begin{abstract}
	
	Ever since its first release in 2009, the Go programming language (Golang) has been well received by software communities. A major reason for its success is the powerful support of library-based development, where a Golang project can be conveniently built on top of other projects by referencing them as libraries. As Golang evolves, it recommends the use of a new library-referencing mode to overcome the limitations of the original one. While these two library modes are incompatible, both are supported by the Golang ecosystem. 
	The heterogeneous use of library-referencing modes across Golang projects has caused numerous dependency management (DM) issues, incurring reference inconsistencies and even build failures.
	Motivated by the problem, we conducted an empirical study to characterize the DM issues, understand their root causes, and examine their fixing solutions.
	Based on our findings, we developed \textsc{Hero}, an automated technique to detect DM issues and suggest proper fixing solutions. We applied \textsc{Hero} to 19,000 popular Golang projects. The results showed that \textsc{Hero} achieved a high detection rate of 98.5\% on a DM issue benchmark and found 2,422 new DM issues in 2,356 popular Golang projects.
	We reported 280 issues, among which 181 (64.6\%) issues have been confirmed, and 160 of them (88.4\%) have been fixed or are under fixing. Almost all the fixes have adopted our fixing suggestions.

\end{abstract}


\section{Introduction}
\label{sec:Introduction}
\sethlcolor{grey}

The Go programming language (Golang) is quickly adopted by software practitioners since its release in 2009~\cite{yasir2019godexpo}.
Like other modern languages, Golang
allows a project to import and reuse functionalities from another Golang project (i.e., library) by simply specifying an \emph{import path}~\cite{Import_path_syntax}.
There are four popular sites hosting Golang projects, namely, Bitbucket~\cite{Bitbucket}, GitHub~\cite{GitHub}, Launchpad~\cite{Launchpad}, and IBM DevOps Services~\cite{IBMDevOpsServices}.
Among them, GitHub hosts nearly 90\% Golang projects (as of June 2020)~\cite{libraries.io.go}. 

While Golang's library-based development boosts its adoption, its \emph{library-referencing mode} has undergone a major change as the language evolves. 
Prior to Golang 1.11, library-referencing was supported by the {\mycode GOPATH} mode. Libraries referenced by a project are fetched using command {\mycode go get}~\cite{Go_get}. 
This mode does not require developers to provide any configuration file.
It works by matching the URLs of the site hosting referenced libraries with the import paths specified by the {\mycode go get} command.
However, it fetches only a library's latest version.
To overcome this restriction, developers use third-party tools such as {\mycode Dep}~\cite{Dep} and {\mycode Glide}~\cite{Glide} to manage different library versions under the \emph{Vendor directory}\footnote{The \emph{Vendor} attribute allows a Golang project to reference a library's different versions and keep them in different folders under a vendor directory.}.
To satisfy developers' need for referencing specific library versions, in August 2018, Golang 1.11 introduced the {\mycode Go Modules} mode, which allows multiple library versions to be referenced by a module using different paths. A \emph{module} comprises a tree of Golang source files with a {\mycode go.mod} configuration file defined in the tree's root directory. The configuration file explicitly specifies the module's dependencies with specific library versions as well as a \emph {module path} by which the module itself can be uniquely referenced by other projects.
The file must be specified according to the \emph{semantic import versioning} (SIV) rules~\cite{Go_Modules}.
For instance, projects whose major versions are \emph{v2} or above should include a version suffix like \hl{``\emph{/v2}''} at the end of their module paths. 

Compared with {\mycode GOPATH}, {\mycode Go Modules} is flexible and allows multiple library versions to coexist in a Golang project~\cite{SIV}.
Developers are suggested to migrate their projects' library-referencing modes from {\mycode GOPATH} to {\mycode Go Modules}. 
However, the migration took a long time. We sampled 20,000 popular Golang projects on GitHub. 
As of June 2020, only 35.9\% projects had migrated to {\mycode Go Modules}, while the rest 64.1\% were still using {\mycode GOPATH}, resulting in the coexistence of two different library-referencing modes. What's more, many projects suffered from various \textit{dependency management} (DM) issues caused by such mixed library-referencing modes. 
Specifically, we made the following three observations:
\begin{itemize}[leftmargin=*, topsep=4pt, itemindent=8pt]
	\item \emph{{\mycode Go Modules} is not backward compatible with {\mycode GOPATH}.} There are two scenarios.
	First, a Golang project can be referenced by its downstream projects. After it migrates to {\mycode Go Modules}, its introduced virtual import paths (with version suffixes) cannot be recognized by downstream projects still in {\mycode GOPATH}. This causes build errors to these projects. 
	Second, a downstream project that has migrated to {\mycode Go Modules} may not find its referenced libraries in {\mycode GOPATH}, or may fetch unintended library versions, due to different import path interpretations by the two modes.

	
	\item \emph{SIV rules can be violated even if a Golang project and its referenced upstream projects both use {\mycode Go Modules}.}
	For instance, a project of major version \emph{v2} may not necessarily include a version suffix at the end of its module path.
	Such violation can be due to developers' misunderstanding or weak SIV rule enforcement (discussed later in Sec II-A).
	They can cause a large number of unresolved library references (``\emph{cannot find package}'' errors) when downstream projects are built.
	
	
	\item \emph{Resolving DM issues for a Golang project 
		requires up-to-date knowledge of its upstream and downstream projects, as well as their possible heterogeneous uses of two library-referencing modes.} However, such information is not provided by the Golang ecosystem to help developers evaluate a solution's impact on other projects. 
	Resolving a DM issue in a project locally without considering the ecosystem in a holistic way can easily cause new issues to its downstream projects. 
\end{itemize}

\begin{figure}[t!]
	\centering
	\includegraphics[width=0.49\textwidth]{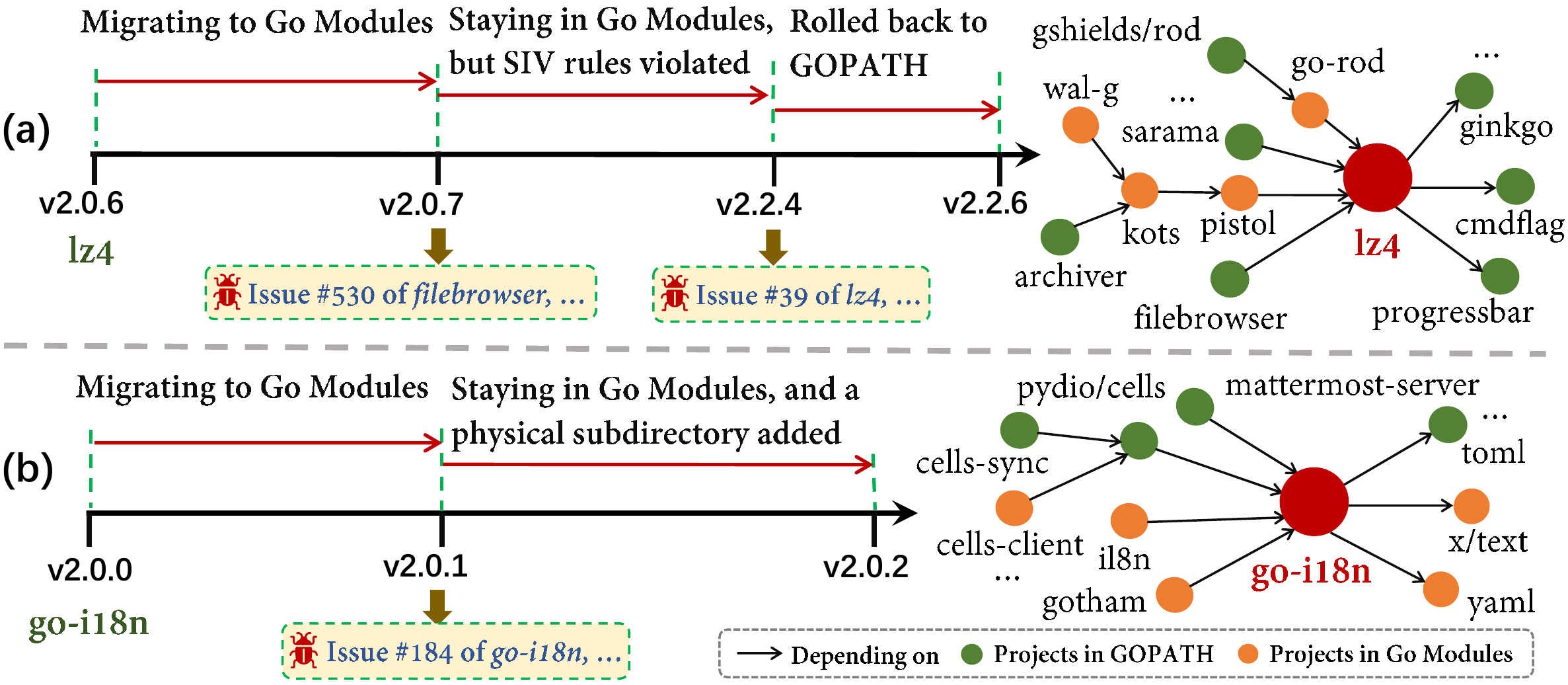}
	\vspace{-5mm}
	\caption{DM issue examples}
	\label{example}
	\vspace{-6mm}
\end{figure}

Figure~\ref{example}(a) shows a DM issue example. 
Project {\mycode lz4}~\cite{pierrec/lz4} migrated to {\mycode Go Modules} in version v2.0.7. 
Following SIV rules, it declared module path \hl{\emph{github.com/pierrec/lz4/v2}} in its {\mycode go.mod} file with version suffix \hl{``\emph{/v2}''}. Although the project can be built successfully after migration, it induced DM issues to downstream projects still in {\mycode GOPATH}, since the latter cannot recognize the version suffix in module paths (e.g., issue \#530 of {\mycode filebrowser}~\cite{Issue530}). 
To fix the problem, {\mycode lz4} released version v2.2.4, 
which was still in {\mycode Go Modules} but removed version suffix \hl{``\emph{/v2}''} from its module path as a workaround. 
This resolved the DM issues in its downstream projects in {\mycode GOPATH}, but induced build errors into its downstream projects that had already migrated to {\mycode Go Modules}, since this solution violated SIV rules (e.g., issue \#39 of {\mycode lz4}~\cite{Issue39}). 
As there is no accurate way to estimate the migration impact to its downstream projects, {\mycode lz4} chose to roll back to {\mycode GOPATH} in v2.2.6 and suspended its migration until its most downstream projects had completed migrations. Such problems are common across Golang projects, imposing unforeseeable risks in mode migration. 
Figure~\ref{example}(b) shows another example from {\mycode go-i18n}~\cite{go-i18n}. 
Its version v2.0.1 followed its upstream projects to use {\mycode Go Modules} for finer library-referencing control. 
However, such change induced at least five DM issues to downstream projects in {\mycode GOPATH} (e.g., issue \#184~\cite{Issue184}) due to their inability to interpret version suffix \hl{``\emph{/v2}''} in {\mycode go-i18n}'s module path. 
To fix the problem, {\mycode go-i18n} v2.0.2's repository provided an additional subdirectory \hl{\emph{go-i18n/v2}} with a copy of implementations to support downstream projects in {\mycode GOPATH}. 
This is a suboptimal solution since it changes the virtual path in {\mycode Go Modules} into a physical one that requires extra maintenance in every project release.
In fact, without a holistic view of all dependencies and the interference between their mixed library-referencing modes, it is hard for developers to find a proper solution to fix DM issues without impacting downstream projects.

Such chaos caused by mixed library-referencing modes is unique to Golang ecosystem, unlike existing dependency conflict issues in Java~\cite{wang2018dependency, wang2019could, huang2020interactive}, JavaScript~\cite{patra2018conflictjs} and Python projects~\cite{wang2watchman}.
Besides, our study of 20,000 Golang projects on GitHub suggests the severity of DM issues caused by the mode migration, since a majority of these projects have chosen to stay with old {\mycode GOPATH}.
To better dig into the problem, we sampled 500 projects from top 1,000 ones, and collected 151 DM issues from the issue trackers for a deeper study of their characteristics and solutions.
We identified three DM issue patterns and summarized eight fixing solutions commonly adopted by developers.
Leveraging these findings, we further developed an automated tool, \textsc{Hero} (\underline{HE}alth diagnosis tool fo\underline{R} the g\underline{O}lang ecosystem), to detect DM issues.
One interesting feature is that \textsc{Hero} can also provide customized fixing suggestions to developers with analyses of potential benefits and consequences incurring to the ecosystem.
To evaluate \textsc{Hero}, we collected 132 real DM issues from top 1,000 Golang projects that were not used in our empirical study and conducted experiments using these issues as a benchmark.
\textsc{Hero} achieved a detection rate of 98.5\% (only missed two cases).
We further applied \textsc{Hero} to the rest 19,000 projects collected from GitHub, and detected 2,422 new DM issues. 
We submitted 280 of them that were associated with the 1,001$^{st}$\textendash2,000$^{th}$ popular projects, and suggested fixing solutions.
Encouragingly, 181 (64.6\%) issues have been confirmed, and 160 (88.4\%) of them have been fixed or under fixing using our suggested solutions.
Such fixes would cause minimal or acceptable impacts to other projects in the ecosystem.
The confirmed issues cover well-known projects, such as {\mycode github/hub}~\cite{github/hub}  
and {\mycode microsoft/presidio}~\cite{microsoft/presidio}, and have promoted 29 projects' migration to {\mycode Go Modules}.

To summarize, in this paper, we made three contributions: 

\begin{itemize}[leftmargin=*, topsep=1pt]
	\item \emph{Originality.} To our best knowledge, we conducted the first empirical study on 20,000 Golang projects to investigate their status of library-referencing mode migration and analyze 151 real DM issues to unveil their characteristics.

	\item \emph{Technique.}
	We developed the \textsc{Hero} tool\footnote{http://www.hero-go.com/} to diagnose dependency management issues for the Golang ecosystem. It can detect DM issues effectively and provide customized fixing suggestions.

	\item \emph{Reproduction package.} We provided a reproduction package on \textsc{Hero} website for future research, which includes: (1) detailed information of the 20k subjects and 151 DM issues studied in our empirical study;
	(2) our benchmark dataset (132 DM issues and subjects used for evaluation);
	
	
\end{itemize}

\section{Background}
\label{sec:Background}

We introduce SIV rules in {\mycode Go Modules} and the concept of module-awareness, to facilitate our later discussions.

\vspace{-2mm}
\subsection{SIV Rules in {\mycode Go Modules}}
\label{sec:SIV}

\begin{figure}[t!]
	\centering
	\includegraphics[width=0.48\textwidth]{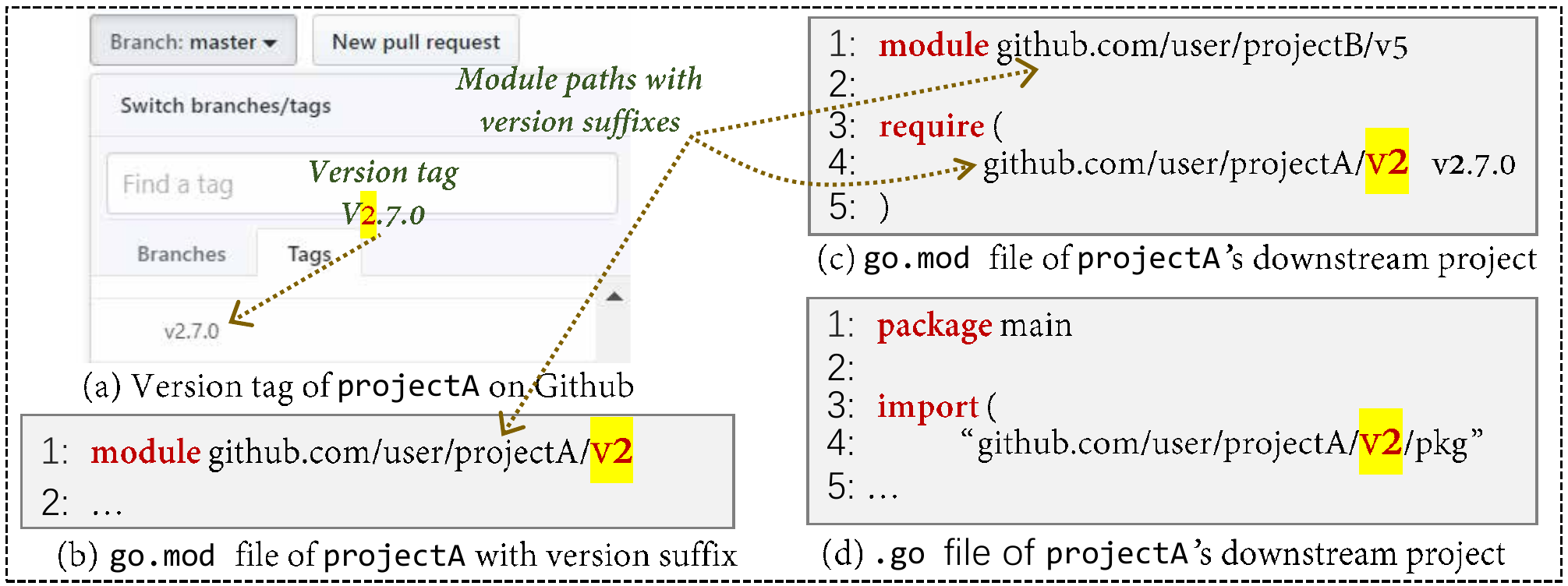}
	\vspace{-1mm}
	\caption{SIV rules in the {\mycode Go Modules} mode}
	\label{Gopath_modules}
\end{figure}

{\mycode Go Modules} introduces SIV to support dependency management of multiple project versions. It has three rules:

\begin{enumerate}[leftmargin=*, topsep=4pt]
	\item Golang projects should follow a semantic versioning format (Semver)\footnote{The Semver format is {\mycode MAJOR}.{\mycode MINOR}.{\mycode PATCH}, where {\mycode MAJOR}, {\mycode MINOR}, and {\mycode PATCH} denote incompatible API changes, backward compatible API changes, and backward compatible bug fixes, respectively (https://semver.org/).}. 
	Figure~\ref{Gopath_modules}(a) gives an example, where {\mycode projectA} tags a release with a semantic version of v2.7.0 on GitHub. 
	
	\item When a project's major version is \emph{v2} or above (denoted as \emph{v2+}), a version suffix like \hl{``\emph{/v2}''} must be included at the end of its module path declared in the {\mycode go.mod} file. 
	As shown in Figure~\ref{Gopath_modules}(b), {\mycode projectA} v2.7.0's module path is \hl{``\emph{github.com/user/projectA/v2}''}. 
	To reference it, downstream projects must declare this path and import it in \emph{require directive} attributes of the {\mycode go.mod} file, as well as in \emph{import directive} attributes of their {\mycode .go} source files. Figures~\ref{Gopath_modules}(c) and (d) give two examples. 
	
	\item If a project's major version is \emph{v0}/\emph{v1}, its version suffix should not be included in its module or import paths.
	
\end{enumerate}

Under these SIV rules in {\mycode Go Modules}, multiple major versions of a library can be separately referenced by different paths.
In contrast, a project in {\mycode GOPATH} can reference only the latest version of a library.

To be more flexible, the official Golang documentation~\cite{Go_Modules} suggests two strategies to release a \emph{v2+} project, namely, \emph{major branch} and \emph{major subdirectory}. 
The former is to update a project's module and import paths to include a version suffix like \hl{``\emph{/v2}''}. 
It is not necessary to physically create a new branch labeled with such a version suffix on the version control system of hosting site. 
The latter is to physically create a subdirectory (e.g., \hl{\emph{projectA/v2}}) with source code and a corresponding {\mycode go.mod} file,
and the corresponding module path must end with a version suffix like \hl{``\emph{/v2}''} accordingly.

As such, module and import paths in the major branch strategy are virtual, but are physical in the major subdirectory strategy.
The latter is sometimes used to provide a transition for downstream projects in {\mycode GOPATH}, as shown in Figure~\ref{example}(b).

\begin{figure}[t!]
	\centering
	\includegraphics[width=0.48\textwidth]{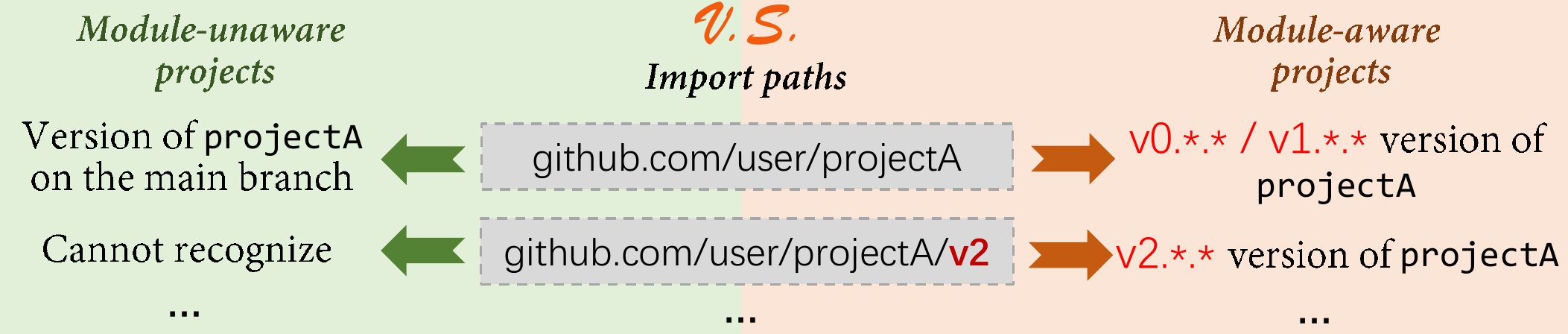}
	\vspace{-1mm}
	\caption{Comparison of module-aware and module-unaware projects} 
	\label{path}
	\vspace{-4mm}
\end{figure}

\sethlcolor{grey}

\subsection{Module-Awareness in Different Golang Versions}

To ease discussion, we refer to the capability of recognizing a virtual path ended with a version suffix like \hl{``\emph{/v2}''} as \emph{module-awareness}. This capability is important for referencing libraries in the Golang ecosystem.

As the migration from {\mycode GOPATH} to {\mycode Go Modules} has immense impact on many Golang projects, it was gradually achieved by multiple Golang versions over two years. 
During migration, ``minimal module compatibility'' was adopted since Golang 1.9.7 in the series 1.9.* and Golang 1.10.3 in the series 1.10.*, which added module-awareness to projects that had not migrated to {\mycode Go Modules}~\cite{minimal}. 
As such, we refer to the versions in range [1.0.1, 1.9.7) $\cup$ [1.10.1, 1.10.3), which manage dependencies in {\mycode GOPATH} without module-awareness, as \emph{legacy Golang versions}. We refer to those in range [1.9.7, 1.10.1) $\cup$ [1.10.3, 1.11.1), which manage dependencies in {\mycode GOPATH} with module-awareness, as \emph{compatible Golang versions}. We refer to those of 1.11.1 or above, which allow projects to adopt either {\mycode GOPATH} or {\mycode Go Modules} and support module-awareness, as \emph{new Golang versions}. We observe that Golang projects in {\mycode GOPATH} often use third-party tools (e.g.,  {\mycode Dep}~\cite{Dep}, {\mycode Glide}~\cite{Glide}, etc.) to help manage dependencies.
Since none of the tools supports ``minimal module compatibility'', their uses actually block module-awareness, messing up library-referencing (e.g., issue \#878 of {\mycode olivere}~\cite{Issue878Elastic} about using {\mycode Dep} and {\mycode glide}, and \#103 of {\mycode migrate}~\cite{Issue103migrate} about using {\mycode govendor}).


Table 1 summarizes module-awareness in different Golang versions. 
Based on this, we give two definitions below:

\begin{table}[]
	\label{Module}
	\centering
	\scriptsize
	\setlength\tabcolsep{2.6pt}     
	\def\arraystretch{1.0}
	
	\caption{Module awareness in different Golang versions}
	\vspace{-1mm}
	\bgroup
	\begin{tabular}{c||c|c|c|c}
		\toprule
		\rowcolor[HTML]{EFEFEF} 
		Category                                                                               & Version range                                                                                  & \begin{tabular}[c]{@{}c@{}}DM\\ mode\end{tabular} & \begin{tabular}[c]{@{}c@{}}Using\\ DM tools\end{tabular} & \begin{tabular}[c]{@{}c@{}}Module\\ awareness\end{tabular} \\ \hline \hline
		&                                                                                                  &                                                        & Y                                                        &                                                            \\ \cline{4-4}
		\multirow{-2}{*}{\begin{tabular}[c]{@{}c@{}}Legacy\\ Golang versions\end{tabular}}     & \multirow{-2}{*}{\begin{tabular}[c]{@{}c@{}}{[}1.0.1, 1.9.7)\\ $\cup$ {[}1.10.1, 1.10.3)\end{tabular}}  & \multirow{-2}{*}{{\mycode GOPATH}}                               & N                                                        & \multirow{-2}{*}{N}                                        \\ \hline
		&                                                                                                  &                                                        & Y                                                       & N                                                          \\ \cline{4-5} 
		\multirow{-2}{*}{\begin{tabular}[c]{@{}c@{}}Compatible\\ Golang versions\end{tabular}} & \multirow{-2}{*}{\begin{tabular}[c]{@{}c@{}}{[}1.9.7, 1.10.1)\\ $\cup$ {[}1.10.3, 1.11.1)\end{tabular}} & \multirow{-2}{*}{{\mycode GOPATH}}                               & N                                                        & Y                                                          \\ \hline
		&                                                                                                  &                                                        & Y                                                        & N                                                          \\ \cline{4-5} 
		&                                                                                                  & \multirow{-2}{*}{{\mycode GOPATH}}                               & N                                                        & Y                                                          \\ \cline{3-5} 
		\multirow{-3}{*}{\begin{tabular}[c]{@{}c@{}}New \\ Golang versions\end{tabular}}       & \multirow{-3}{*}{$\geq$ 1.11.1}                                                            & {\mycode Go Modules}                                             & \textendash                                                        & Y                                                          \\ \bottomrule
		\multicolumn{5}{l}{DM stands for \underline{d}ependency \underline{m}anagement. ``\textendash'' means ``not applicable''.}\\	
	\end{tabular}
	\egroup
	\vspace{-4mm}
\end{table}

\textbf{Definition 1 (Module-aware project):}
A project is \textit{module-aware} if and only if it uses a compatible or new Golang version and does not use any DM tool.


\textbf{Definition 2 (Module-unaware project):} 
A project is \textit{module-unaware} if and only if it uses a legacy Golang version, or it uses a compatible or new Golang version with a DM tool.

Figure~\ref{path} shows how module-aware and module-unaware projects differ in parsing an import path with or without a \emph{v2+} version suffix. 
For an import path like \hl{\emph{github.com/user/projectA}}, a module-aware project could reference a specific version \emph{v0.$\ast$.$\ast$} or \emph{v1.$\ast$.$\ast$} of {\mycode projectA} under \emph{v2} (latest version under \emph{v2}, by default), while a module-unaware project would reference the version on {\mycode projectA}'s main branch (typically the latest version).
For an import path like \hl{\emph{github.com/user/projectA/v2}}, the former could reference a specific version \emph{v2.$\ast$.$\ast$} of {\mycode projectA} (latest version under \emph{v3}, by default), while the latter would fail to recognize it. 

According to the above background knowledge, we formally define the DM issues occurred in Golang projects as follows:

\textbf{Definition 3 (Dependency management (DM) issue):} 
If an issue is caused by the different interpretations between module-aware and module-unaware projects or violating SIV rules by {\mycode Go Modules} projects, we refer to it as a \textit{DM issue} in Golang ecosystem.

A project suffers from a DM issue may fetch the unintended versions of its libraries, or may not find its referenced libraries.

\section{Empirical Study}
\label{sec:Empirical Study}
We empirically study the characteristics of DM issues and the scale of these issues arising from the varying degrees of module-awareness in different Golang versions. 
We aim to answer the following three research questions:

\begin{itemize}[leftmargin=*, topsep=1pt]
	\item \textbf{RQ1 (Scale of Module-Awareness)}: \textit{What is the status quo of library-referencing mode migration for projects in the Golang ecosystem? To what extent are they module-aware?}
	
	\item \textbf{RQ2 (Issue Types and Causes)}: \textit{What are common types of DM issues? What are their root causes?}
	
	\item \textbf{RQ3 (Fixing Solutions)}: \textit{What are common practices for fixing DM issues? How do they affect the ecosystem?}
\end{itemize}

To answer RQ1, we collected top 20,000 popular and active open-source Golang projects from GitHub to study their migration status.
To answer RQ2/3, we randomly selected 500 subjects (denoted as $subject Set_1$) from top 1,000 of our collected projects.
We then collected real DM issues from these projects plus some additional ones.
To dig into these issues, we manually analyzed their issue descriptions, developers' discussions, code commits, and the Golang official documentation.
Note that the rest 500 projects (denoted as $subject Set_2$) in top 1,000 of our collected projects were not used in RQ2/3. They are used to evaluate our DM issue detection technique later (Sec~\ref{sec:Effectiveness}). Below we present our data collection procedure and study results in detail.

\subsection{Data Collection}
\label{sec:Data Collection}
\textbf{Step 1: Collecting Golang projects.}
We collected top 20,000 popular and active Golang projects from GitHub, which hosts over 90\% Golang ones. 
A project's \emph{popularity} is decided by its star counts, and \emph{activeness} is decided based on whether 50+ code commits exist in its repository since Jan 2020.

\begin{figure}[t!]
	\centering
	\includegraphics[width=0.48\textwidth]{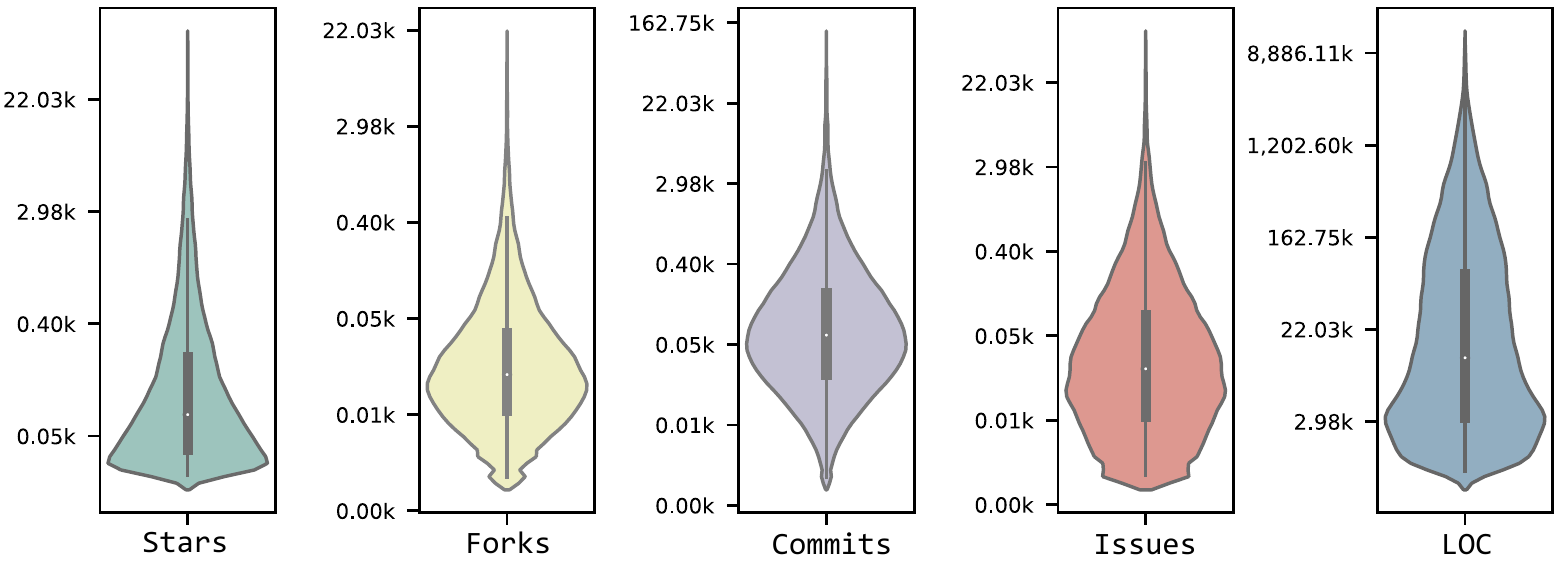}
	\vspace{-1mm}
	\caption{Statistics of collected 20,000 Golang projects (log scale)}
	\label{Subjects}
\end{figure}

Figure~\ref{Subjects} shows these projects' demographics. They are: (1) popular (60.3\% having 100+ stars or forks), (2) well-maintained (on average having 339 code commits and 136 issues), and (3) large-sized (on average having 72.3 KLOC).
We used these projects for RQ1. 


\textbf{Step 2: Collecting DM issues.} 
For the 500 projects in $subject Set_1$, after filtering the ones that have no issue trackers or code repositories, we considered the remaining 484 projects as subjects.
We then added to the seed subjects Golang's official project {\mycode golang/go}~\cite{golang/go} and two most popular dependency management tools {\mycode Dep}~\cite{Dep} and {\mycode Glide}~\cite{Glide}, for better studying DM issues from the perspective of the ecosystem.
In total, we obtained 487 projects for RQ2/3. 

As these projects contain many issue reports, we filtered using keywords ``go modules'' and ``go.mod'' (case insensitive) to locate potential DM issues for manual analysis (``go.mod'' configuration file is a notable new feature in the {\mycode Go Modules} mode). 
Keyword ``go modules'' returned 1,342 issue reports, and ``go.mod'' returned 2,421 ones.
We merged overlapping reports and then removed noise. First, we excluded issue reports that did not discuss DM issues
(e.g., issue \#5559~\cite{Issue5559} of project {\mycode gogs}~\cite{gogs} only documented developers' plan to migrate to {\mycode Go Modules}).
Second, we excluded issue reports that discussed nothing about root causes of DM issues.

Three co-authors cross-checked all collected issue reports, and finally obtained a collection of 151 well-documented DM issues, which involves 127 Golang projects.
They contain sufficient details for studying RQ2/3.


\subsection{RQ1: Scale of Module-Awareness}
\label{sec:RQ1}
We analyze the scale of module-awareness as below:

\begin{itemize}[leftmargin=*, topsep=4pt]
	\item For all 20,000 projects, we counted the number of projects that have migrated to {\mycode Go Modules} by checking whether  {\mycode go.mod} files exist in their latest versions' repositories.
	

	\item For projects that have migrated to {\mycode Go Modules}, we checked whether their major version numbers of latest releases are \emph{v2+}.
	If so, we further checked their adopted strategies (i.e., major branch/subdirectory) in the code repositories.
	
	
	\item For projects still in {\mycode GOPATH}, we checked whether they use third-party tools to manage dependencies by the presence of their configuration files.
	For example, using the {\mycode Dep}~\cite{Dep} or {\mycode Glide}~\cite{Glide} tool requires a {\mycode Gopkg.toml} or {\mycode glide.yaml} configuration file, respectively.
	
	
	
\end{itemize}

\begin{figure}[t!]
	\centering
	\includegraphics[width=0.5\textwidth]{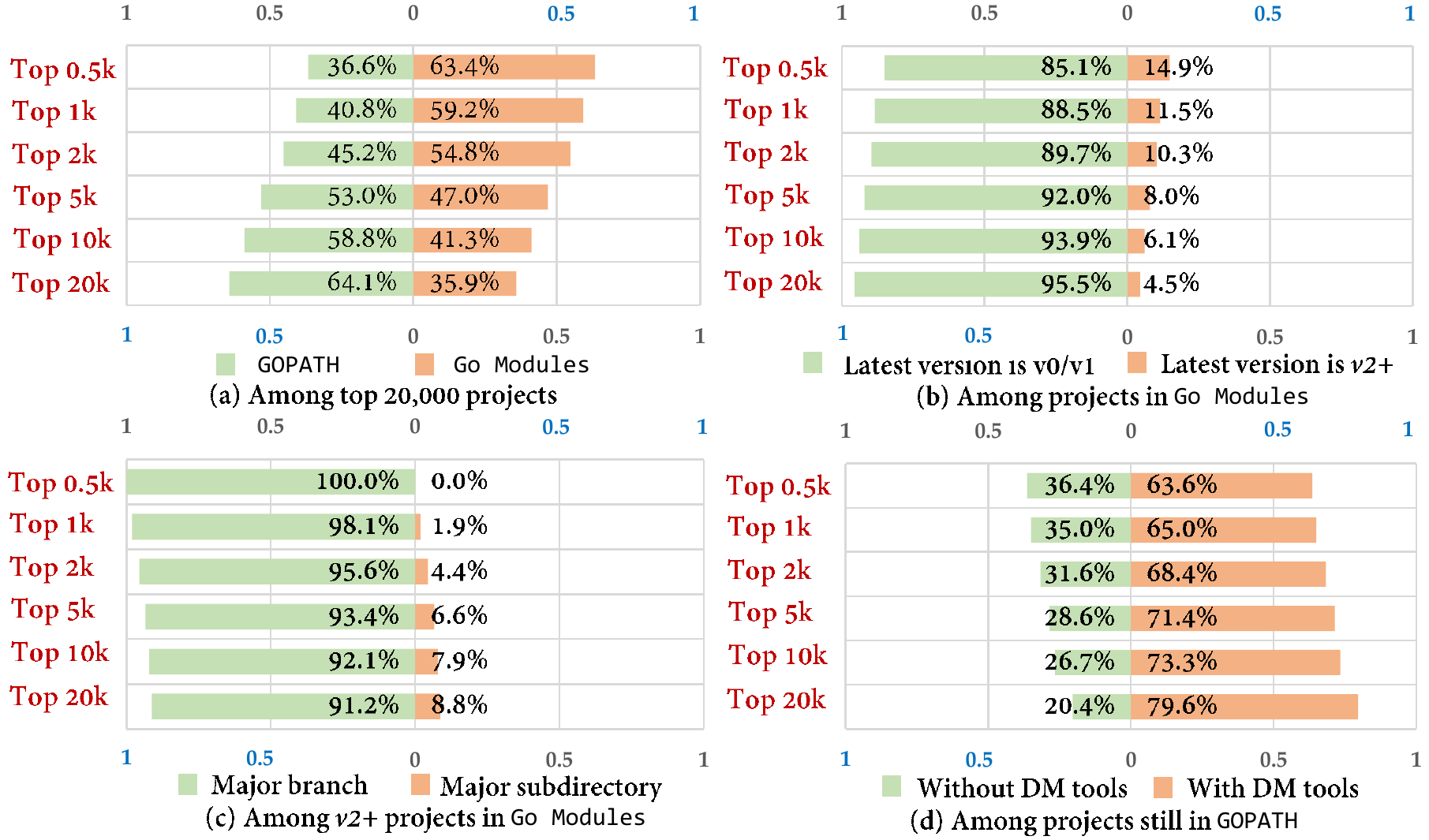}
	\vspace{-4mm}
	\caption{Investigation statistics for RQ1}
	\label{RQ1}
	\vspace{-5mm}
\end{figure}

\textbf{\emph{Results.}} 
Figure~\ref{RQ1} shows analysis results.
To see trends, we divided all projects into six (overlapping) groups based on their popularities: top 500, 1k, 5k, 10k, and 20k (1k = 1,000).

From Figure~\ref{RQ1}(a), we see that the proportion of {\mycode Go Modules} migrations increases with the popularity of projects. 
This suggests that migrating to {\mycode Go Modules} is a good practice in the ecosystem. 
Still, 64.1\% projects are in {\mycode GOPATH} despite that two years have passed since {\mycode Go Modules} came into being.

Figures~\ref{RQ1}(b) and~\ref{RQ1}(c) show that only 4.5\% projects that have migrated to {\mycode Go Modules} released \emph{v2+} versions (i.e., most ones are still in \emph{v0/v1}), and 91.2\% \emph{v2+} versions were managed by the major branch strategy.
This suggests that the vast majority of \emph{v2+} projects should be referenced by virtual module paths ended with version suffixes like \hl{``\emph{/v2}''}.
Then they are likely to induce build failures in module-unaware downstream projects.
Besides, for the rest 95.5\% projects whose major versions are \emph{v0/v1}, DM issues can easily occur when they are updated to \emph{v2+} versions in future.

Figure 4(d) shows that 79.6\% projects in {\mycode GOPATH} use third-party tools to manage dependencies.
As aforementioned, this will block module-awareness for projects that adopt compatible Golang versions.
Therefore, at least 10,205 of top 20,000 Golang projects (51.0\%) are module-unaware.

\textbf{\emph{Challenges of migration.}} 
Our findings may explain why many Golang projects stay with {\mycode GOPATH}.
We also investigate how developers consider this problem from projects still in {\mycode GOPATH}. We focused on the {\mycode GOPATH} part (36.6\%) of top 500 out of the 20,000 projects (Fig.~\ref{RQ1}(a)), and analyzed their issue reports that discuss migration to study reasons for holding the migration.
We obtained 52 issue reports specifically discussing unsuccessful migration, and observed three common reasons:
\begin{itemize}[leftmargin=*, topsep=0pt]
	
	\item \emph{\textbf{Existing versioning scheme incompatible with SIV rules in {\mycode Go Modules} (27/52).}}
	Some projects have their own versioning schemes, different from SIV rules in {\mycode Go Modules}.
	To avoid incompatibility (e.g., issue \#328 of {\mycode go-tools}~\cite{Issue328}), developers chose to stay with {\mycode GOPATH}.
	
	\item \emph{\textbf{Third-party DM tools hindering the migration plan (15/52).}}
	Some projects heavily rely on third-party tools for dependency management.
	As the tools do not work with {\mycode Go Modules}, developers chose to live with the tools instead of migration (e.g., issue \#61~\cite{Issue61} of {\mycode uuid}).
	
	\item \emph{\textbf{Causing problems to downstream projects in {\mycode GOPATH} (10/52).}} 
	Many projects are still in {\mycode GOPATH}, inconvenient to reference upstream projects in {\mycode Go Modules}.
	For continuous support for downstream projects, developers chose to stay with {\mycode GOPATH} (e.g., issue \#103~\cite{Issue103migrate} of {\mycode migrate}).

	
\end{itemize}

Due to these challenges, 
we conjecture that {\mycode GOPATH} and {\mycode Go Modules} can co-exist for a long time.
This suggests the inevitability of DM issues in the Golang ecosystem and motivates us to study their characteristics and fixing solutions.

\tcbset{
	colback=green!5!white, colframe=black,
	notitle, 
	width={\linewidth+1pt},
	top=0pt,
	left=1pt,
	right=1pt,
	bottom=0pt,
	toprule=0.5pt,
	titlerule=1pt,
	bottomrule=0.5pt,
	leftrule=0.5pt,
	rightrule=0.5pt,
	after skip=6pt,}
\vspace{4pt}
\noindent\begin{tcolorbox}
	\noindent\emph{\textbf{Answer to RQ1:}} \emph{
		Golang projects face challenges in migrating to {\mycode Go Modules}. Up till June 2020, only 35.9\% of top 20,000 projects on GitHub have migrated to {\mycode Go Modules}, and
		at least 51.0\% of top 20,000 projects are module-unaware.
		The two library-referencing modes may co-exist for a long time in the ecosystem. 
	}
	
\end{tcolorbox}

\vspace{0.2em}

\subsection{RQ2: Issue Types and Root Causes}
\label{sec: Issue Types and Causes}

We observed three common types of DM issues in collected issue reports. 
Below we introduce them and analyze their root causes with examples.


\textbf{\emph{Type A.}} \emph{DM issues can occur when projects in {\mycode GOPATH} depend on projects in {\mycode Go Modules} (41/151=27.2\%).}
The former are typically module-unaware. Build errors can occur when such projects directly or transitively depend on the latter but cannot recognize their virtual paths with version suffixes, e.g., issue \#1017 of {\mycode glide}~\cite{Issue1017glide}.

Among 41 \emph{Type A} issues, 35 occurred in module-unaware projects when they upgraded upstream dependencies whose newer versions introduced virtual import paths.
This shows that version upgrades of libraries in {\mycode Go Modules} can impose threats to their module-unaware downstream projects and developers should estimate such threats before upgrading.
The rest 6 issues occurred when introducing new upstream projects that transitively depend on virtual import paths.

\textbf{\emph{Type B.}} \emph{DM issues can occur when projects in {\mycode Go Modules} depend on projects in {\mycode GOPATH} (40/151=26.5\%).}
There are two cases. The first (\emph{Type B.1}) is due to the different import path interpretations between {\mycode GOPATH} and {\mycode Go Modules}, and the second (\emph{Type B.2}) is due to the interference of \emph{Vendor} attribute in {\mycode GOPATH}.


\emph{\textbf{Type B.1} (16/40).} Let project $P_A$ in {\mycode Go Modules} depend on project $P_B$ in {\mycode GOPATH}, and $P_B$ further depend on $P_C$ in {\mycode Go Modules} with import path \hl{\emph{github.com/user/PC}}.
Suppose that $P_C$ has released a \emph{v2+} version with the major branch strategy.
From $P_B$'s perspective, it interprets the import path as $P_C$'s latest version (i.e., \emph{v2+} version on $P_C$'s main branch).
However, in $P_A$'s build environment, the import path is interpreted as a \emph{v0}/\emph{v1} version of $P_C$ (no version suffix in the path).
As a result, $P_A$ fails to fetch $P_C$'s correct version and can encounter errors when building with $P_B$.

\textit{Type B.1} issues are difficult to notice, and can easily cause build errors.
For example, issue \#47246 of {\mycode cockroach}~\cite{Issue47246} reported that a client project in {\mycode Go Modules} depends on {\mycode cockroach} v19.5.2 in {\mycode GOPATH}, and {\mycode cockroach} further depends on project {\mycode apd}~\cite{apd} in {\mycode Go Modules} (with a \emph{v2+} version).
Although {\mycode cockroach} itself correctly referenced {\mycode apd} v2.0.0 (latest version) by interpreting import path \hl{\emph{github.com/cockroachdb/apd}}, the client project instead fetched {\mycode apd} v1.1.0 based on its interpretation of this import path.
As a result, the client project's building failed due to missing an important field (not in {\mycode apd} v1.1.0 but in v2.0.0).

\emph{\textbf{Type B.2} (24/40).} 
Let project $P_A$ in {\mycode Go Modules} depend on project $P_B$ in {\mycode GOPATH}, and $P_B$ further depend on project $P_C$, which is managed in $P_B$'s \emph{Vendor} directory.
A \emph{Vendor} directory is a major feature of {\mycode GOPATH}, which localizes the maintenance of remote dependencies' specific versions.
We note that $P_A$ references $P_C$ by import path \hl{\emph{github.com/user/PC}} declared in $P_B$'s source files rather than from $P_B$'s \emph{Vendor} directory.
Although the build may work for the time being, $P_A$ can fail to fetch $P_C$ if $P_C$ is deleted or moved to another repository (e.g., renaming).
Even if the fetching is successful, the version on $P_C$'s hosting site could be different from the one in $P_B$'s \emph{Vendor} directory, causing potential build errors due to the inconsistency.


Such situations often occur, since there are essentially two versions of a library at two different sites and their consistency is not guaranteed.
We witnessed a \textit{Type B.2} issue in project {\mycode moby}~\cite{moby}, which has received 57.6k stars on GitHub and ranked the third in popularity.
To support its large number of downstream projects still in {\mycode GOPATH}, {\mycode moby} has not migrated to {\mycode Go Modules}.
Its issue \#39302~\cite{Issue39302} reported that {\mycode moby} referenced project {\mycode logrus}~\cite{logrus} from its \emph{Vendor} directory, and {\mycode logrus} had been relocated from \hl{\emph{github.com/\textbf{S}irupsen/logrus}} to \hl{\emph{github.com/\textbf{s}irupsen/logrus}} (case sensitive) on GitHub.
This incurred DM issues to many of {\mycode moby}'s downstream projects in {\mycode Go Modules} (e.g., issues \#127 of {\mycode testcontainers}~\cite{Issue127} and issue \#2 of {\mycode shnorky}~\cite{Issue2}), as they could not fetch {\mycode logrus} by the import path in {\mycode moby}'s source files.

\textbf{\emph{Type C.}} \emph{DM issues can occur when projects in {\mycode Go Modules} depend on projects also in {\mycode Go Modules} but not following SIV rules (70/151=46.4\%).}
We identified three types of SIV rule violations that caused build failures to downstream projects:
(1) lacking version suffixes like \hl{``\emph{/v2}''} in module paths or import paths, although the versions of concerned projects are \emph{v2+} (37/70) (e.g., issue \#1355~\cite{Issue1355} of {\mycode iris});
(2) version tags not following the {\mycode MAJOR}.{\mycode MINOR}.{\mycode PATCH} format (18/70) (e.g., issue \#1848~\cite{Issue1848} of {\mycode gobgp});
(3) module paths in {\mycode go.mod} files are inconsistent with URLs associated with concerned projects on their hosting sites (15/70) (e.g., issue \#9~\cite{Issue9} of {\mycode jwplayer}).

While downstream projects can encounter build failures, the projects violating SIV rules do not produce warnings or errors themselves when building.
Currently, there is no diagnosis technique to detect the three SIV rule violation types, or mechanism to enforce SIV rules, as discussed in issues \#1355 of {\mycode iris}~\cite{Issue1355} and \#32695 of {\mycode golang/go}~\cite{Issue32695} (by {\mycode lz4}'s~\cite{Issuelz4} users).
As a result, projects violating SIV rules can ``safely'' stay in the Golang ecosystem, despite the unexpected consequences to their downstream projects.
Regarding such risk, {\mycode lz4}'s developers commented its severity on issue \#32695 that
\emph{``we need to fix this issue and figure out how big the crater it brings to the ecosystem.''}

\tcbset{
	colback=green!5!white, colframe=black,
	notitle, 
	width={\linewidth+1pt},
	top=0pt,
	left=1pt,
	right=1pt,
	bottom=0pt,
	toprule=0.5pt,
	titlerule=1pt,
	bottomrule=0.5pt,
	leftrule=0.5pt,
	rightrule=0.5pt,
	after skip=6pt,}
\vspace{4pt}
\noindent\begin{tcolorbox}
	\noindent\emph{\textbf{Answer to RQ2:}} \emph{
		DM issues commonly occur due to heterogeneous uses of {\mycode GOPATH} and {\mycode Go Modules}. Their manifestations can be summarized into three types and there are two common root causes: (1) {\mycode GOPATH} and {\mycode Go Modules} interpret import paths in different ways, and (2) SIV rules are not strictly enforced in the Golang ecosystem.	
	}
\end{tcolorbox}
\vspace{0.2em}
\subsection{RQ3: Fixing Solutions}
\label{sec:Fixing Solutions}

Out of the 151 DM issues, 144 issues have fixing patches or fixing plans that developers have agreed on. We studied them and observed eight common fixing solutions, which demonstrate different trade-offs. 


\textbf{\emph{Solution 1:}} \emph{Projects in {\mycode GOPATH} migrate to {\mycode Go Modules} (22/144=15.3\%).} 
Migrating from {\mycode GOPATH} to {\mycode Go Modules} can help fix \emph{Type A} issues, since these issues are caused by projects still in {\mycode GOPATH}, which are unable to recognize import paths with version suffixes.
For example, in issue \#454~\cite{Issue454}, {\mycode redis}~\cite{redis} migrated to {\mycode Go Modules}, but its downstream project {\mycode benthos} was still in {\mycode GOPATH}.
Then, {\mycode benthos} was suggested to migrate to {\mycode Go Modules} to avoid build errors.
This solved {\mycode benthos}'s problem, but caused incompatibility to {\mycode benthos}'s module-unaware downstream projects.
As a result, new \emph{Type A} issues (e.g., issue \#232~\cite{Issue232}) arose.

\textbf{\emph{Solution 2:}} \emph{Projects in {\mycode Go Modules} roll back to {\mycode GOPATH} (13/144=9.0\%).} 
Some projects rolled back to {\mycode GOPATH}  after migrating to {\mycode Go Modules} for fixing \emph{Types A} and \emph{C} issues.
For example, in issue \#61~\cite{Issue61} (\emph{Type A}), project {\mycode uuid}'s [46] migration to {\mycode Go Modules} broke the building of many downstream projects in {\mycode GOPATH}.
As a compromise, {\mycode uuid} rolled back to {\mycode GOPATH}, waiting for downstream projects to migrate first.
In issue \#663~\cite{Issue663} (\emph{Type C}), {\mycode gopsutil} and its downstream projects were all in {\mycode Go Modules}, but {\mycode gopsutil} violated SIV rules (lacking a version suffix in its module path of \emph{v2+} release), causing build errors to downstream projects.
As such, {\mycode gopsutil} chose to roll back to {\mycode GOPATH} to make downstream projects work again.
This solution solves the problem, but hinders the migration status of the ecosystem.

\textbf{\emph{Solution 3:}} \emph{Changing the strategy of releasing v2+ projects in {\mycode Go Modules} from major branch to subdirectory (6/144=4.2\%).} 
It helps resolve \emph{Type A} issues, where module-unaware projects cannot recognize virtual import paths for \emph{v2+} libraries in {\mycode Go Modules}.
The new strategy creates physical paths by code clone, so that libraries can be referenced by module-unaware projects.
However, this is just a workaround and needs extra maintenance in subsequent releases (e.g., issue of {\mycode go-i18n}~\cite{Issue184} as discussed in Sec~\ref{sec:Introduction}).


\textbf{\emph{Solution 4:}} \emph{Maintaining v2+ libraries in {\mycode Go Modules} in downstream projects' Vendor directories rather than referencing them by virtual import paths (6/144=4.2\%).}
Similar to \emph{solution 3}, this solution also helps resolve \emph{Type A} issues.
By making a copy of libraries in downstream projects' repositories, it avoids fetching the libraries by virtual import paths. For example, in issue \#141~\cite{Issue141}, {\mycode radix}~\cite{radix} refused to use the major subdirectory strategy for its \emph{v2+} project release in {\mycode Go Modules}. Its downstream projects had to make a copy of {\mycode radix}'s code in their \emph{Vendor} directories, which requires extra maintenance and potentially cause \emph{Type B.2} issues in future.

\begin{figure}[t!]
	\centering
	\includegraphics[width=0.48\textwidth]{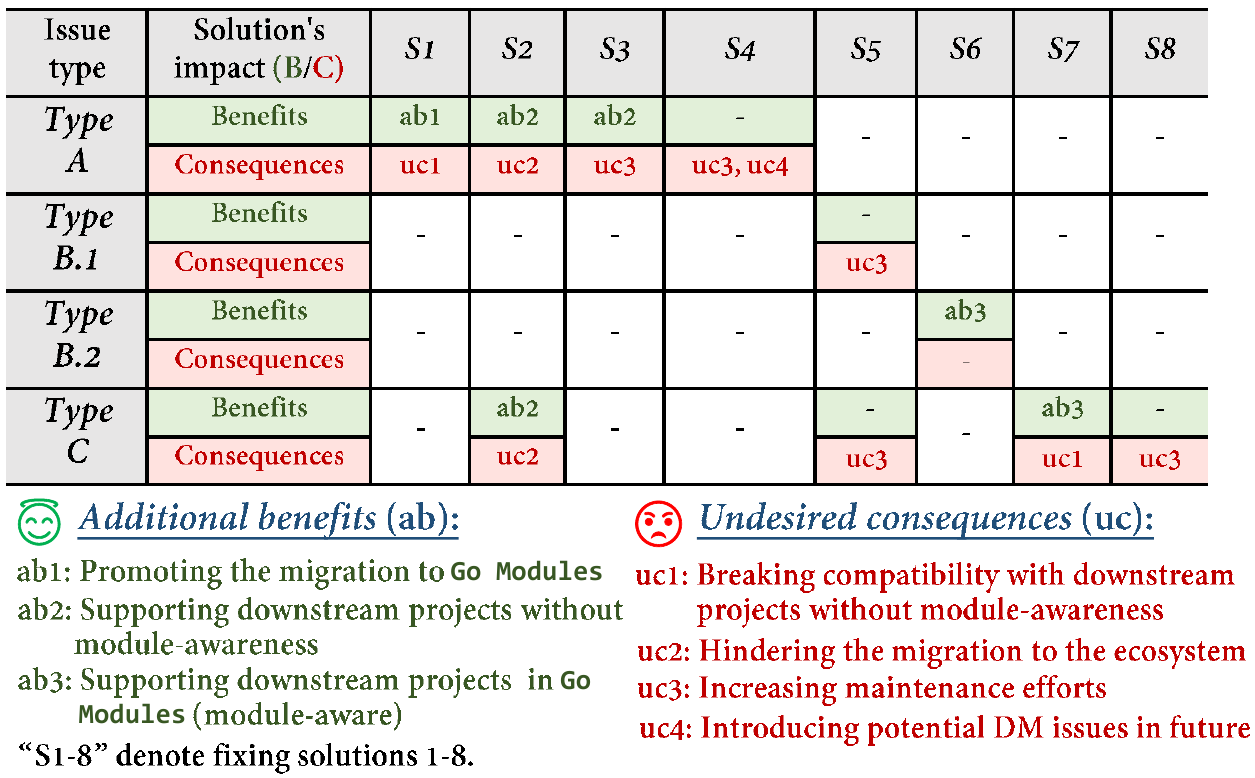}
	\vspace{-3mm}
	\caption{Benefits and consequences of the eight fixing solutions}
	\label{benefitcost}
	\vspace{-6mm}
\end{figure}

\textbf{\emph{Solution 5:}} \emph{Using a replace directive with version information to avoid using import paths in referencing libraries (16/144=11.1\%).} 
It addresses \emph{Types B.1} (problematic import path interpretations) and \emph{Type C} (import path violating SIV rules) issues.
For example, in issue \#12~\cite{Issue12}, a client project used a directive to replace the original import path: \hl{replace github.com/andrewstuart/goq => astuart.co/goq v1.0.0}, to reference its expected project {\mycode goq}'s version~\cite{goq}.
However, this would make developers no longer able to use the {\mycode go get} command to automatically fetch upgraded libraries.

\textbf{\emph{Solution 6:}} \emph{Updating import paths for libraries that have changed their repositories (24/144=16.7\%).} 
It fixes \emph{Type B.2} issues, where libraries in a project's \emph{Vendor} directory may be inconsistent with the ones referenced by their import paths.
It updates import paths to help a project's downstream projects in {\mycode Go Modules} fetch consistent library versions.
For example, in issue \#429~\cite{Issue429}, {\mycode go-cloud} 
managed library {\mycode etcd} in its \emph{Vendor} directory, {\mycode etcd} later changed its hosting repository from \hl{\emph{github.com/coreos/etcd}} to \hl{\emph{go.etcd.io/etcd}}.
To fix build errors for its downstream projects in {\mycode Go Modules},  {\mycode go-cloud} updated {\mycode etcd}'s import path to the latest one for the consistency.
This fixes the issue and benefits all affected downstream projects without impacting others in the ecosystem.

\textbf{\emph{Solution 7:}} \emph{Projects in {\mycode Go Modules} fix configuration items to strictly follow SIV rules (47/144=32.6\%).} 
Projects that have migrated to {\mycode Go Modules} are suggested to follow Golang's official guidelines on SIV rules to fix their induced \emph{Type C} issues.
For example, in \#1149~\cite{Issue1149}, project {\mycode redis}~\cite{redis} added a version suffix \hl{``\emph{/v7}''} at the end of its module path to follow SIV rules.
However, we noticed that while the issues are fixed, the project's downstream projects in {\mycode GOPATH} may be impacted (unable to recognize the version suffixes, e.g., issue \#1151~\cite{Issue1151} reported for {\mycode redis}).

\textbf{\emph{Solution 8:}} \emph{Using a hash commit ID for a specific version to replace a problematic version number in library referencing (10/144=6.9\%).} 
It fixes \emph{Type C} issues, where some projects in {\mycode Go Modules} violate SIV rules in version numbers and cause build errors to downstream projects that are also in {\mycode Go Modules}.
It avoids referencing problematic version numbers, by a \emph{require directive} with a specific hash commit ID. 
For example, in issue \#6048~\cite{Issue6048}, one of {\mycode prometheus}'s downstream projects in {\mycode Go Modules} chose to use directive \hl{require github.com/prometheus/prometheus 43acd0e} to reference its expected version v2.12.0.
Similar to \emph{Solution 5}, this solution would also make developers unable to automatically fetch upgraded libraries using command {\mycode go get}.

As summarized in Figure~\ref{benefitcost}, these solutions fix their targeted DM issues, but at the same time they may bring additional benefits ($ab1$\textendash$ab3$) or undesired consequences ($uc1$\textendash$uc4$).
When there are multiple fixing solutions for a specific DM issue, developers are suggested to carefully consider the relevant dependencies and minimize the impact on other projects in the ecosystem, by weighing consequences against benefits.

\tcbset{
	colback=green!5!white, colframe=black,
	notitle, 
	width={\linewidth+1pt},
	top=0pt,
	left=1pt,
	right=1pt,
	bottom=0pt,
	toprule=0.5pt,
	titlerule=1pt,
	bottomrule=0.5pt,
	leftrule=0.5pt,
	rightrule=0.5pt,
	after skip=6pt,}
\vspace{4pt}
\noindent\begin{tcolorbox}
	\normalsize
	\noindent\emph{\textbf{Answer to RQ3:}} \emph{We observed eight common fixing solutions for DM issues, covering 95.4\% of the studied issues.
		Most solutions could affect other projects in the ecosystem.
		When fixing a DM issue, developers should find a tradeoff between the benefits and the possible consequences. 
	}
\end{tcolorbox}
\vspace{0.2em}

\section{\textsc{Hero}: DM Issue Diagnosis}
\label{sec:approach}
Our empirical study reveals the prevalence of DM issues in the Golang ecosystem due to the chaotic use of {\mycode GOPATH} and {\mycode Go Modules} in different projects.  
This motivates us to develop a tool, named \textsc{Hero}, to help automatically detect DM issues and provide customized fixing solutions.
\textsc{Hero} works in two steps. It first extracts dependencies among Golang projects and their library-referencing modes and then diagnoses DM issues in these projects based on our observed issue types and root causes (RQ2). It further provides customized fixing suggestions leveraging the findings in RQ3.
\textsc{Hero} can analyze a single Golang project or monitor the heterogeneous use of the two library-referencing modes in the Golang ecosystem.
Below we explain how \textsc{Hero} models project dependencies and detects DM issues.

\subsection{Constructing Dependency Model}
We first build a dependency model for the Golang project under analysis. We formally define the model below. 

\textbf{Definition 3 (Dependency model):}
The dependency model $\mathcal{D}(P_v)$ for version $v$ of a project $P$ is a 3-tuple $(Pr, Ds, Us)$:

\begin{itemize}[leftmargin=*, topsep=4pt]
	\item $Pr = (ip, md, t, vd)$ records the information of the \textbf{current project},  where $ip$ and $md$ are $P_v$'s declared module path (for $P_v$ to be referenced by downstream projects) and library-referencing mode ({\mycode GOPATH} or {\mycode Go Modules}), respectively.
	If $P_v$ is in {\mycode GOPATH}, fields $t$ and $vd$ denote whether $P_v$ depends on any DM tool ({\mycode yes} or {\mycode no}), and a collection of import paths (set of URLs) referencing those upstream libraries that are maintained in $P_v$'s \emph{Vendor} directory but cannot be found in the repositories pointed to by URLs (e.g., due to removal or renaming), respectively.
	Otherwise, the two fields are set to {\mycode no} and {\mycode null}, respectively.

	\item $Ds = \{dp_1, dp_2, \cdots, dp_n\}$ is a collection of $P_v$'s \textbf{downstream projects} $dp_i$, where $dp_i = (v_i, ip_i, md_i, t_i)$.
	Field $v_i$ denotes $dp_i$'s latest version number.
	Fields $ip_i$, $md_i$, and $t_i$ denote this version's import path, library-referencing mode, and whether any DM tool is used, respectively.

	\item $Us = \{up_1, up_2, \cdots, up_n\}$ is a collection of $P_v$'s \textbf{upstream projects} $up_i$, where $up_i = (v_i, ip_i, md_i, S_i, I_i)$.
	The fields $ip_i$ and $md_i$ denote $v_i$'s import path and library-referencing mode, respectively.
	If $P_v$ is in {\mycode Go Modules}, field $v_i$ denotes $up_i$'s specific version declared in $P_v$'s configuration file.
	Otherwise (i.e., when $P_v$ is in {\mycode GOPATH}), $v_i$ denotes $up_i$'s latest version number.
	If $up_i$ is a \emph{v2+} project in {\mycode Go Modules}, field $S_i$ denotes whether it is released by the major branch strategy ({\mycode yes} or {\mycode no}), implying whether $ip_i$ is a virtual import path.
	If both projects $up_i$ and $P_v$ are in {\mycode Go Modules}, field $I_i$ denotes whether $up_i$ is transitively introduced into $P_v$ by any project in {\mycode GOPATH} ({\mycode yes} or {\mycode no}).
	Otherwise, the two fields are set to {\mycode null} and {\mycode no}, respectively.

\end{itemize}

\begin{figure*}[t!]
	\centering
	\includegraphics[width=0.94\textwidth]{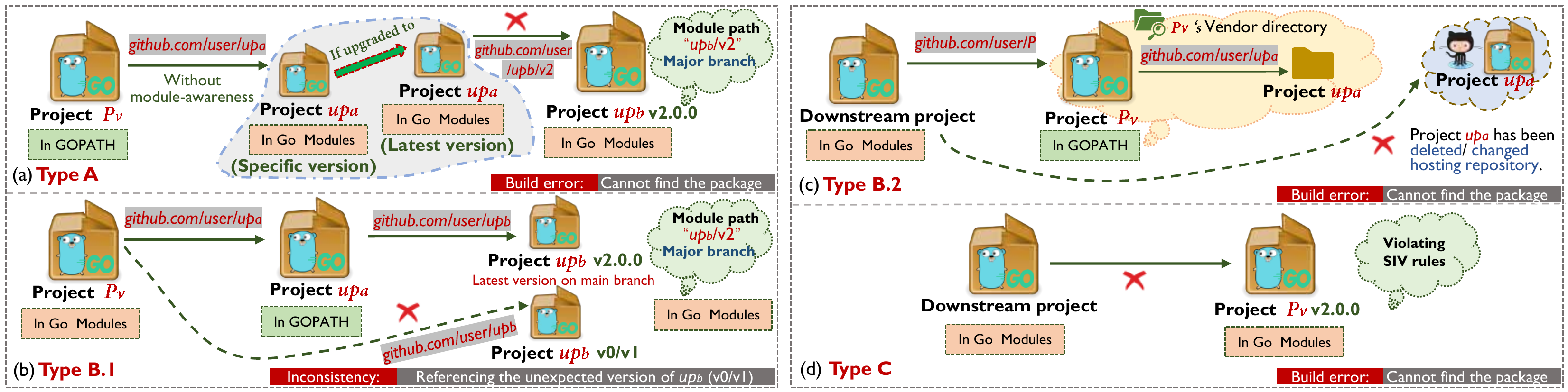}
	\caption{Three types of DM issues \textsc{Hero} detects}
	\label{type}
\end{figure*}

\begin{figure*}[t!]
	\centering
	\includegraphics[width=1\textwidth]{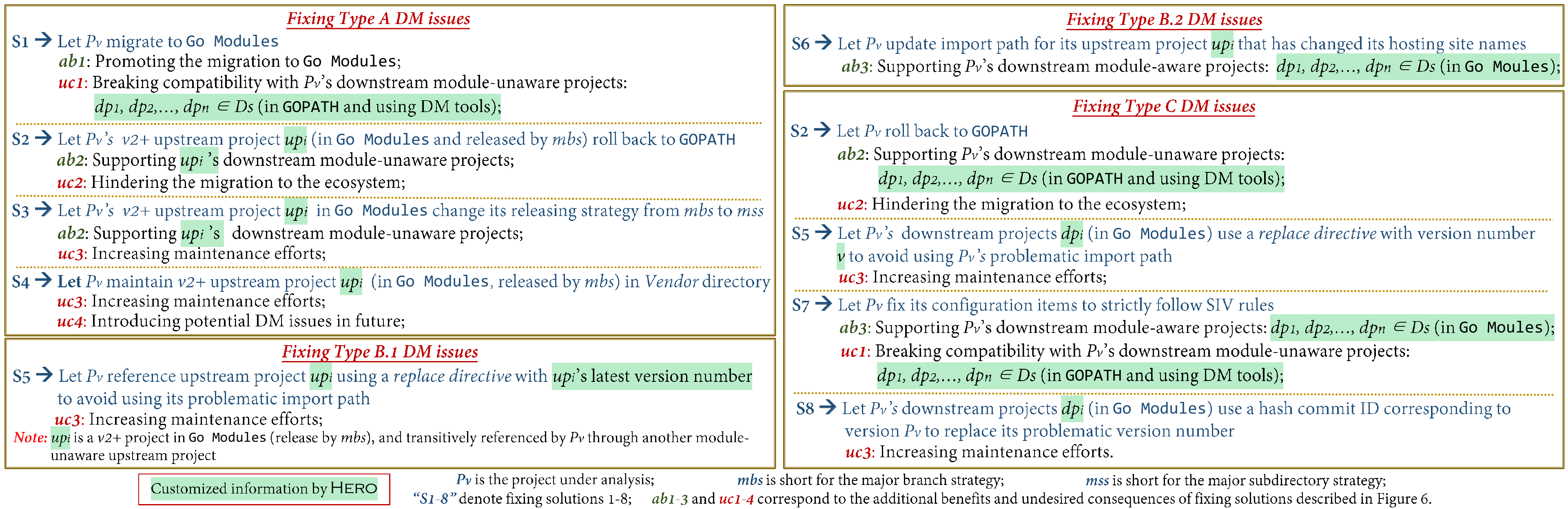}
	\caption{Templates of customized fixing suggestions for three types of DM issues}
	\label{Template}
\end{figure*}

We explain how to obtain these field values, taking GitHub (the most popular Golang project hosting site) for example:

\textbf{\emph{Step 1: Collecting $Pr$ information.}}
Leveraging GitHub's REST API ``\emph{repository\_url}''~\cite{REST}, \textsc{Hero} queries with $P_v$'s repository name to obtain its import path $ip$ and library-referencing mode $md$ by checking if a {\mycode go.mod} file exists in its repository.
If $P_v$ is in {\mycode GOPATH}, \textsc{Hero} decides field $t$ by checking whether any DM tool's configuration file exists. 
Field $vd$ is decided by parsing $P_v$'s source files to collect import paths for libraries maintained in the \emph{Vendor} directory, and querying via the  ``\emph{repository\_url}'' API with the collected import paths to check whether the corresponding libraries have been deleted or relocated (e.g., by {\mycode HTTP 404: Not Found} errors~\cite{REST}).

\textbf{\emph{Step 2: Collecting $Ds$ information.}} Leveraging GitHub's REST API ``\emph{code\_search\_url}''~\cite{REST}, \textsc{Hero} queries with $P_v$'s repository name to check which projects depend on it.
This information is from the \emph{require directives} of a project's {\mycode go.mod} file, \emph{import directives} of its source files, or a DM tool's configuration file.
Each found project corresponds to an item $dp_i$ in the collection $Ds$.
Note that \textsc{Hero} collects the latest version $v_i$ for $dp_i$, and decides its associated import path $ip_i$, library-referencing mode $md_i$ (by checking whether $P_v$'s repository name is declared in its {\mycode go.mod} file), and field $t_i$ (by checking whether its DM tools' configuration file exists), respectively.
These collected downstream projects depend on $P_v$ and can also reference its earlier versions. 

\textbf{\emph{Step 3: Collecting $Us$ information.}}
Project $P_v$'s upstream projects information is collected in two ways, depending on the library referencing mode of the project:
\begin{itemize}[leftmargin=*, topsep=4pt]
	\item $P_v$ in {\mycode Go Modules}: \textsc{Hero} collects $P_v$'s upstream projects $up_i$ with fields $ip_i$ and $v_i$ by parsing its {\mycode go.mod} file, which configures a project's direct and transitive dependencies with import paths and specific version numbers.
	\textsc{Hero} identifies $up_i$'s library-referencing mode $md_i$ by checking whether a {\mycode go.mod} file exists in its repository via GitHub's ``\emph{repository\_url}'' API.
	If $up_i$ is a \emph{v2+} project in {\mycode Go Modules}, \textsc{Hero} identifies its release strategy $S_i$ by checking whether a subdirectory like \hl{\emph{``$up_i$/v2''}} exists.
	For projects transitively introduced into $P_v$ by any project in {\mycode GOPATH}, Golang's build tool automatically marks them with a ``\emph{//indirect}'' comment at the end of their module paths in $P_v$'s {\mycode go.mod} file~\cite{Gomod}, with which \textsc{Hero} decides $I_i$.

	\item $P_v$ in {\mycode GOPATH}: \textsc{Hero} collects $P_v$'s direct dependencies $up_i$ with import paths $ip_i$ from its source files. With the import paths, \textsc{Hero} leverages GitHub's ``\emph{repository\_url}'' API to look into these dependencies' repositories to collect their latest versions, from which it decides the corresponding version numbers $v_i$ and library-referencing modes $md_i$.
	Then \textsc{Hero} recursively collects the information of $P_v$'s transitive dependencies declared in {\mycode go.mod} or sources files in concerned repositories, and identifies version numbers, import paths, library-referencing modes in a similar way.

\end{itemize}


\subsection{Diagnosing DM Issues}
\vspace{-1mm}
The dependency model built by \textsc{Hero} contains sufficient information for detecting DM issues and suggesting solutions.


\textbf{Detecting DM issues.} 
Our study disclosed that most DM issues caused build errors, already observable.
Thus, \textsc{Hero} focuses on detecting DM issues that have not yet manifested, but would probably happen when the concerned projects have their upstream or downstream projects upgraded.
Due to page limit, we explain scenarios for which \textsc{Hero} reports issues in this paper with algorithm details on our website.


\emph{\textbf{Type A.}} Figure~\ref{type}(a) shows a scenario, where a module-unaware project $P_v$ references a specific version of its upstream project $up_a$ in {\mycode Go Modules}.
This version is older than $up_a$'s latest version, which newly introduces another upstream project $up_b$ in {\mycode Go Modules} with a \emph{v2+} version released using the major branch strategy.
Build errors do not occur in $P_v$ when it references $up_a$'s old version. 
However, if $P_v$ updates $up_a$ to reference the latest version, it will not be able to recognize $up_b$'s virtual import path. When seeing such a possibility, \textsc{Hero} reports a warning of \emph{Type A} issue for $P_v$.

\emph{\textbf{Type B.1.}} Figure~\ref{type}(b) shows a scenario, where project $P_v$ in {\mycode Go Modules} transitively references a \emph{v2+} upstream project $up_b$ in {\mycode Go Modules} (released by the major branch strategy) through another module-unaware project $up_a$ in {\mycode GOPATH}.
Since {\mycode GOPATH} and {\mycode Go Modules} interpret import paths differently, $up_a$ would use $up_b$'s latest version (e.g., v2.0.0), while $P_v$ would use $up_b$'s old \emph{v0/v1} version, causing inconsistencies.
Thus, \textsc{Hero} reports a warning of \emph{Type B.1} issue for $P_v$.

\emph{\textbf{Type B.2.}} Figure~\ref{type}(c) shows a scenario, where project $P_v$ in {\mycode GOPATH} references an upstream project $up_a$ maintained only in its \emph{Vendor} directory (i.e., $up_a$ has already been deleted or relocated). 
No build errors occur when $P_v$ has no downstream projects in {\mycode Go Modules}. However, if $P_v$ has such downstream projects, the latter would fetch $up_a$ via its import path (i.e., hosting repository) rather than from $P_v$'s \emph{Vendor} directory, causing build errors due to failing to fetch $up_a$.
Thus, \textsc{Hero} reports a warning of \emph{Type B.2} issue for $P_v$.

\emph{\textbf{Type C.}} Figure~\ref{type}(d) shows a scenario, where project $P_v$ in {\mycode Go Modules} violates SIV rules (as discussed in Sec~\ref{sec: Issue Types and Causes}). 
The violation may not introduce build errors when $P_v$ has no downstream projects in {\mycode Go Modules}.
However, build errors would occur if such projects exist in future.
Thus, \textsc{Hero} reports a warning of \emph{Type C} issue for $P_v$.

\textbf{\emph{Customized fixing suggestions.}} Our empirical study has identified applicable fixing solutions for each issue type (Figure~\ref{benefitcost}). We summarize the impacts of these solutions as templates in Figure~\ref{Template}.
For each detected DM issue, \textsc{Hero} suggests all applicable solutions to developers by customizing the template with potential impact analysis based on the associated dependency model.

\section{Evaluation}
\label{sec:Evaluation}

We study two research questions in our evaluation of \textsc{Hero}:

\begin{itemize}[leftmargin=*, topsep=1pt]
	\item \textbf{RQ4 (Effectiveness)}: \textit{How effective is \textsc{Hero} in detecting DM issues for Golang projects?}
	
	\item \textbf{RQ5 (Usefulness)}: \textit{Can \textsc{Hero} detect new DM issues for real-world Golang projects and assist the developers in fixing the detected issues?}
\end{itemize}

For RQ4, we conducted experiments using the 132 DM issues from the 500 Golang projects in $sujectSet2$.
Note that none of them overlap with those issues used in our empirical study.
Specifically, we constructed a benchmark dataset containing the 132 DM issues and their project versions for evaluating whether \textsc{Hero} can detect these issues in the buggy versions or predict them in earlier versions.
It is worth mentioning that issue-fixing versions are not necessarily issue-free, since new DM issues can be introduced after fixing as we have discussed earlier.

For RQ5, we applied \textsc{Hero} to the rest 19,000 of the top 20,000 Golang projects (i.e., excluding 500 used for RQs 2\textendash3 and 500 used for RQ4).
We reported the detected issues together with root cause analyses and fixing suggestions to respective developers.
In our issue reports, we also highlighted the preferred solutions based on their impact on other projects.


\vspace{-1mm}
\subsection{RQ4: Effectiveness}
\label{sec:Effectiveness}
\vspace{-1mm}

\emph{\textbf{Experimental setup.}} 
The benchmark dataset contains 38 \emph{Type A} (28.8\%), 15 \emph{Type B.1} (11.3\%), 28 \emph{Type B.2} (21.2\%), and 51 \emph{Type C} (38.6\%) DM issues.
We collected their corresponding project versions to evaluate \textsc{Hero}'s capability of detecting or predicting DM issues:

\begin{itemize}[leftmargin=*, topsep=4pt]
	\item \emph{Type A:} These issues occurred when module-unaware projects in {\mycode GOPATH} referenced \emph{v2+} dependencies in {\mycode Go Modules} by virtual import paths.
	Since issue occurrences would already cause build errors, we ran \textsc{Hero} on the previous project versions where such issues had not occurred.

	\item \emph{Type B.1:} These issues occurred when projects in {\mycode Go Modules} referenced dependencies in {\mycode GOPATH}, with different import path interpretations to \emph{v2+} projects released by the major branch strategy. The inconsistency may not lead to immediate build errors or functional failures, but is indeed risky.
	Thus, we ran \textsc{Hero} on the current project versions to check whether it can detect potential issues.

	\item \emph{Types B.2 and C:} The former occurred when the dependencies maintained in the current projects' \emph{Vendor} directories were deleted or relocated remotely. The latter occurred when the current projects in {\mycode Go Modules} violated SIV rules. In both cases, the current projects would not have symptoms like build errors, but their downstream projects in {\mycode Go Modules} would when referencing them in future.
	Thus, we ran \textsc{Hero} on current project versions to check whether it can detect potential issues.
\end{itemize}


\begin{table}[]
	\label{effec}
	\scriptsize
	\centering
	\setlength\tabcolsep{4.6pt}     
	\def\arraystretch{0.97}
	\caption{\textsc{Hero}'s effectiveness on DM issue detection}
	\vspace{-1mm}
	\bgroup
	\begin{tabular}{c||c|c|c|c|c}
		\toprule
		\rowcolor[HTML]{EFEFEF} 
		\diagbox{Result}{Type} & \emph{Type A} & \emph{Type B.1} & \emph{Type B.2} & \emph{Type C} & \textbf{Summary} \\ \hline \hline
		\cellcolor[HTML]{EFEFEF}\emph{Ground truth}          & 38     & 15  & 28   & 51     & 132     \\ \hline
		\cellcolor[HTML]{EFEFEF}\emph{Detected}  & 36     & 15  & 28    & 51     & 130     \\ \hline
		\cellcolor[HTML]{EFEFEF}\emph{Missed} & 2      & 0  & 0  & 0    & 2       \\ \hline
		\cellcolor[HTML]{EFEFEF}\emph{Detection rate}      & 94.7\% & 100\% & 100\% & 100\%  & 98.5\%  \\ \bottomrule
	\end{tabular}
	\egroup
	\vspace{-4mm}
\end{table}

\emph{\textbf{Results.}} Table~2 shows our experiment results.
\textsc{Hero} reported a total of 130 DM issues (all true positives), covering 98.5\% issues in the benchmark dataset.
\textsc{Hero} achieved such a high detection rate because it constructs a dependency model that captures all necessary information on the characteristics of common DM issues.
The only two missing issues are of \emph{Type~A}.
\textsc{Hero} failed to detect them due to its conservative nature in identifying module-aware projects in {\mycode GOPATH} without using any DM tools.
We note that precisely deciding module-awareness requires checking a project's local build environment to know whether it adopts a compatible Golang version.
Currently, \textsc{Hero} does not support such checking.

\vspace{-1mm}
\subsection{RQ5: Usefulness}

\begin{table*}[h!t!]
	\definecolor{Status1}{HTML}{458B74}
	\definecolor{Status2}{HTML}{76EEC6}
	\definecolor{Status3}{HTML}{B4EEB4}
	\definecolor{Status4}{HTML}{CDC673}
	\definecolor{Status5}{HTML}{f5f5f5}
	\setlength\fboxsep{1.6pt}
	\normalsize
	\centering
	\label{table2}
	\caption{Statistics of 280 DM issues reported by \textsc{Hero}}
	\vspace{-3mm}
	\def\arraystretch{1.15}
	\begin{center}
		\resizebox{\textwidth}{!}{
			\begin{tabular}{|p{34pt}|p{970pt}|} 
				\hline
				\multicolumn{1}{|c|}{\textbf{Type}} & \multicolumn{1}{c|}{\textbf{Issue reports (Issue report ID, Project name)}}\\ \hline
				\multirow{6}{*}{\emph{\textbf{Type A}}} & 
				\colorbox{Status1}{\color{white} \#2922,kiali{\color{red}$\spadesuit$};} 
				\colorbox{Status1}{\color{white} \#345,go-carbon{\color{red}$\spadesuit$};} 
				\colorbox{Status1}{\color{white} \#21,gowebsocket{\color{red}$\spadesuit$};} 
				\colorbox{Status1}{\color{white} \#10,nging{\color{red}$\spadesuit$};} 
				\colorbox{Status1}{\color{white} \#8,Hands-On-SE{\color{red}$\spadesuit$};} 
				\colorbox{Status1}{\color{white} \#53,kafka-proxy{\color{red}$\spadesuit$};} 
				\colorbox{Status1}{\color{white} \#456,istio-operator{\color{red}$\spadesuit$};} 
				\colorbox{Status1}{\color{white} \#48,gke-managed-certs{\color{red}$\spadesuit$};} 
				\colorbox{Status1}{\color{white} \#23,render{\color{red}$\spadesuit$};} 
				\colorbox{Status1}{\color{white} \#1068,amazon-ecs-cli{\color{red}$\spadesuit$};} 
				\colorbox{Status2}{\#2488,amazon-ecs-agent{\color{red}$\spadesuit$};}		
				\colorbox{Status2}{\#1647,postgres-operator{\color{red}$\spadesuit$};} 
				\colorbox{Status2}{\#1852,metrictank{\color{red}$\spadesuit$};} 
				\colorbox{Status2}{\#249,cells{\color{red}$\spadesuit$};} 			
				\colorbox{Status2}{\#141,pupernetes{\color{red}$\spadesuit$};} 
				\colorbox{Status2}{\#3840,teleport{\color{red}$\spadesuit$};} 
				\colorbox{Status2}{\#315,fathom{\color{red}$\spadesuit$};} 		
				\colorbox{Status2}{\#222,balena-engine{\color{red}$\spadesuit$};} 
				\colorbox{Status2}{\#491,go-vite{\color{red}$\spadesuit$};} 
				\colorbox{Status2}{\#180,standup-raven{\color{red}$\spadesuit$};} 
				\colorbox{Status2}{\#1008,factomd{\color{red}$\spadesuit$};} 		
				\colorbox{Status2}{\#11,webkubectl{\color{red}$\spadesuit$};} 
				\colorbox{Status2}{\#115,terway{\color{red}$\spadesuit$};}
				\colorbox{Status2}{\#1020,quorum{\color{red}$\spadesuit$};} 
				\colorbox{Status2}{\#44,gh-polls{\color{red}$\spadesuit$};}
				\colorbox{Status2}{\#51,core{\color{red}$\spadesuit$};} 
				\colorbox{Status2}{\#106,integram{\color{red}$\spadesuit$};} 
				\colorbox{Status2}{\#50036,cockroach{\color{red}$\spadesuit$};} 
				\colorbox{Status2}{\#93,kubergrunt{\color{red}$\spadesuit$};} 
				\colorbox{Status2}{\#25105,origin;} 	
				\colorbox{Status2}{\#4835,trafficcontrol;} 
				\colorbox{Status3}{\#519,jaeger-client-go;} 
				\colorbox{Status3}{\#288,RedisShake;} 
				\colorbox{Status3}{\#2786,runtime;} 
				\colorbox{Status3}{\#342,presidio;} 
				\colorbox{Status3}{\#18383,snapd;} 
				\colorbox{Status3}{\#475,pgweb;} 
				\colorbox{Status3}{\#1741,heketi;} 
				\colorbox{Status3}{\#52,sqoop;} 
				\colorbox{Status5}{\#94,acyl;} 
				\colorbox{Status5}{\#385,dns;} 
				\colorbox{Status5}{\#729,bitrise;} 
				\colorbox{Status5}{\#18,kube-iptables;}  
				\colorbox{Status5}{\#3460,minishift;} 	
				\colorbox{Status5}{\#787,veneur;} 
				\colorbox{Status5}{\#77,git-chglog;} 
				\colorbox{Status5}{\#447,mu;} 
				\colorbox{Status5}{\#1545,faas;} 
				\colorbox{Status5}{\#330,arena;} 
				\colorbox{Status5}{\#609,fossa-cli;} 
				\colorbox{Status5}{\#178,vearch;} 
				\colorbox{Status5}{\#1,tepleton;} 
				\colorbox{Status5}{\#277,redis-operator;} 
				\colorbox{Status5}{\#1640,openstorage;} 
				\colorbox{Status5}{\#97,manifest-tool;} 
				\colorbox{Status5}{\#82,wave;} 
				\colorbox{Status5}{\#2962,swarmkit;} 
				\colorbox{Status5}{\#72,k8s-rescheduler;} 
				\colorbox{Status5}{\#741,service-broker-azure;} 
				\colorbox{Status5}{\#1007,appsody;} 
				\colorbox{Status5}{\#13,nginx-clickhouse;} 
				\colorbox{Status5}{\#94,acyl;} 
				\colorbox{Status5}{\#534,kubernikus;} 
				\colorbox{Status5}{\#125,core;} 
				\colorbox{Status5}{\#319,operator-marketplace;} 
				\colorbox{Status5}{\#974,GoSublime;} 
				\colorbox{Status5}{\#713,functions;} 
				\colorbox{Status5}{\#1293,ansible-service-broker;} 
				\colorbox{Status5}{\#658,stork;}
				\colorbox{Status5}{\#21,aliyun-jaeger;}
				\colorbox{Status5}{\#209,boleto-api;}
				\colorbox{Status5}{\#411,postgres\_exporter;}
				\colorbox{Status5}{\#4323,bk-cmdb;}
				\colorbox{Status5}{\#129,gitkube;}
				\colorbox{Status5}{\#3016,pouch;}
				\\
				\hline
				
				\multirow{3}{*}{\textbf{\emph{Type B.1}}} & 
				\colorbox{Status1}{\color{white} \#1411,signalfx-agent;} 
				\colorbox{Status1}{\color{white} \#2,findgs;}
				\colorbox{Status1}{\color{white} \#151,block-explorer;}
				\colorbox{Status1}{\color{white} \#284,watchman;}
				\colorbox{Status1}{\color{white} \#256,flamingo-commerce;}
				\colorbox{Status1}{\color{white} \#3,scifgif;}
				\colorbox{Status1}{\color{white} \#12,ntci;}
				\colorbox{Status1}{\color{white} \#488,benthos;}
				\colorbox{Status1}{\color{white} \#182,go-geom;}
				\colorbox{Status1}{\color{white} \#5366,cds;}
				\colorbox{Status1}{\color{white} \#55,vault-pki-backend;}
				\colorbox{Status2}{\#1,foxtrot;}
				\colorbox{Status2}{\#1220,weave;}
				\colorbox{Status2}{\#663,yorc;}
				\colorbox{Status2}{\#1186,blockatlas;}
				\colorbox{Status2}{\#3843,weave;}
				\colorbox{Status3}{\#37,dataframe-go;}
				\colorbox{Status3}{\#295,serial-vault;}
				\colorbox{Status4}{\#3970,sensu-go;}
				\colorbox{Status5}{\#208,bosun}
				\colorbox{Status5}{\#12,vaultenvporter-go}
				\colorbox{Status5}{\#12,stashvision}
				\colorbox{Status5}{\#136,dsk;}		    
				\colorbox{Status5}{\#71,isopod;}
				\colorbox{Status5}{\#719,sops;}
				\colorbox{Status5}{\#48,awsu;}  
				\colorbox{Status5}{\#82,konfigadm;}
				\colorbox{Status5}{\#23,terraform-pingaccess;}
				\colorbox{Status5}{\#170,rabbitmq\_exporter;}
				\colorbox{Status5}{\#21,pivot;}
				\\
				\hline
				
				\multirow{4}{*}{\textbf{\emph{Type B.2}}} & 
				\colorbox{Status1}{\color{white} \#114,tomato;}
				\colorbox{Status1}{\color{white} \#20,kube-cluster;}
				\colorbox{Status1}{\color{white} \#1,ovpn-tool;}
				\colorbox{Status1}{\color{white} \#1,cache;}
				\colorbox{Status1}{\color{white} \#10,rankdb;}
				\colorbox{Status1}{\color{white} \#4,go-workshops;}
				\colorbox{Status1}{\color{white} \#1306,neo-go;}
				\colorbox{Status1}{\color{white} \#2,chat;}
				\colorbox{Status1}{\color{white} \#232,saferwall;}
				\colorbox{Status1}{\color{white} \#7,hcunit;}
				\colorbox{Status1}{\color{white} \#49,examples;}
				\colorbox{Status1}{\color{white} \#499,cost-model;}
				\colorbox{Status1}{\color{white} \#1,universal-adapter;}
				\colorbox{Status1}{\color{white} \#9215,kyma;}
				\colorbox{Status1}{\color{white} \#12,rboot;}
				\colorbox{Status1}{\color{white} \#1,video-stream;}
				\colorbox{Status1}{\color{white} \#347,server;}
				\colorbox{Status1}{\color{white} \#50,CPU-Pooler;}
				\colorbox{Status2}{\#76,honeyaws;} 
				\colorbox{Status2}{\#190,envman;}
				\colorbox{Status2}{\#104,service-mesh;}
				\colorbox{Status2}{\#1512,skygear-server;}
				\colorbox{Status2}{\#69,go-scm;}
				\colorbox{Status2}{\#2401,paas-cf;}
				\colorbox{Status2}{\#37,aur-out-of-date;}
				\colorbox{Status2}{\#7978,telegraf;}
				\colorbox{Status2}{\#36,rai;}
				\colorbox{Status3}{\#2395,kubernetes-client;}
				\colorbox{Status3}{\#234,dcs-bios;}
				\colorbox{Status3}{\#985,assetto-server;}
				\colorbox{Status3}{\#678,louketo-proxy;}
				\colorbox{Status4}{\#770,terraform-libvirt;}
				\colorbox{Status5}{\#1,subs;}
				\colorbox{Status5}{\#732,bitrise;}
				\colorbox{Status5}{\#31,logrus\_influxdb;}
				\colorbox{Status5}{\#63,albiondata-client;}
				\colorbox{Status5}{\#61,mlmodelscope;}
				\colorbox{Status5}{\#17,dns;}
				\colorbox{Status5}{\#107,multiaddr;}
				\colorbox{Status5}{\#83,remp;}
				\colorbox{Status5}{\#438,snmpcollector;}
				\colorbox{Status5}{\#1,goDistributedCron;}
				\colorbox{Status5}{\#4,field-services;}
				\colorbox{Status5}{\#27,chrly;}
				\colorbox{Status5}{\#27,amanar;}
				\colorbox{Status5}{\#20,sailfish;}
				\colorbox{Status5}{\#15,pike;}
				\colorbox{Status5}{\#143,training;}
				\colorbox{Status5}{\#29,airflow-on-k8s;}
				\colorbox{Status5}{\#17,telegraf-lotus;}
				\\
				\hline
				
				\multirow{8}{*}{\textbf{\emph{Type C}}} &
				\colorbox{Status1}{\color{white} \#34,memory-calculator;}
				\colorbox{Status1}{\color{white} \#54,gormt;}
				\colorbox{Status1}{\color{white} \#162,gocron;}
				\colorbox{Status1}{\color{white} \#8,generic;}
				\colorbox{Status1}{\color{white} \#26,go-sessions;}
				\colorbox{Status1}{\color{white} \#13,keystore-go;}
				\colorbox{Status1}{\color{white} \#49,go-sdk;}
				\colorbox{Status1}{\color{white} \#21,gokiteconnect;}
				\colorbox{Status1}{\color{white} \#2517,hub;}
				\colorbox{Status1}{\color{white} \#265,cameradar;}
				\colorbox{Status1}{\color{white} \#309,server;}
				\colorbox{Status1}{\color{white} \#1638,micro;}
				\colorbox{Status1}{\color{white} \#317,marketstore;}	
				\colorbox{Status1}{\color{white} \#28,media-sort;}
				\colorbox{Status1}{\color{white} \#114,mmock;}
				\colorbox{Status1}{\color{white} \#833,chain33;}
				\colorbox{Status1}{\color{white} \#3,artifex;}
				\colorbox{Status1}{\color{white} \#17,accounting;}
				\colorbox{Status1}{\color{white} \#29,checkmail;}
				\colorbox{Status1}{\color{white} \#7138,jx;}
				\colorbox{Status1}{\color{white} \#5,goDoH;}
				\colorbox{Status1}{\color{white} \#77,gin-admin;}
				\colorbox{Status1}{\color{white} \#158,gosparkpost;}
				\colorbox{Status1}{\color{white} \#4,lsleases;}
				\colorbox{Status1}{\color{white} \#11,bhugo;}
				\colorbox{Status1}{\color{white} \#13,mcwss;}
				\colorbox{Status1}{\color{white} \#15,grpc-proxy;}
				\colorbox{Status1}{\color{white} \#5,wifi-password-qr;}
				\colorbox{Status1}{\color{white} \#2,transcoder;}
				\colorbox{Status1}{\color{white} \#2,pipe-to-me;}
				\colorbox{Status1}{\color{white} \#42,restruct;}
				\colorbox{Status1}{\color{white} \#141,gots;}
				\colorbox{Status1}{\color{white} \#23,huego;}
				\colorbox{Status1}{\color{white} \#8,math-engine;}
				\colorbox{Status1}{\color{white} \#6,iso9660;}
				\colorbox{Status1}{\color{white} \#6,raft-badger;}
				\colorbox{Status1}{\color{white} \#27,tenus;}
				\colorbox{Status1}{\color{white} \#27,go-bitcoind;}
				\colorbox{Status1}{\color{white} \#7,gotime;}
				\colorbox{Status1}{\color{white} \#22,ADtoLDAP;}
				\colorbox{Status1}{\color{white} \#80,lenses-go;}
				\colorbox{Status1}{\color{white} \#113,STNS;}
				\colorbox{Status1}{\color{white} \#504,multus-cni;}
				\colorbox{Status1}{\color{white} \#90,tank;}
				\colorbox{Status1}{\color{white} \#4118,git-lfs;}
				\colorbox{Status1}{\color{white} \#203,vale;}
				\colorbox{Status1}{\color{white} \#25,echo-session;}
				\colorbox{Status1}{\color{white} \#118,mmark;}
				\colorbox{Status1}{\color{white} \#481,chirpstack;}
				\colorbox{Status1}{\color{white} \#1255,ceph-csi;}
				\colorbox{Status1}{\color{white} \#284,aliyun-cli;}
				\colorbox{Status1}{\color{white} \#5268,singularity;}
				\colorbox{Status1}{\color{white} \#6306,provider-google;}
				\colorbox{Status1}{\color{white} \#933,cli;}
				\colorbox{Status1}{\color{white} \#2305,felix;}
				\colorbox{Status1}{\color{white} \#501,aws-nuke;}
				\colorbox{Status1}{\color{white} \#2126,calicoctl;}
				\colorbox{Status1}{\color{white} \#91,goblin;}
				\colorbox{Status1}{\color{white} \#3,sparkzstd;}	
				\colorbox{Status1}{\color{white} \#121,email;}
				\colorbox{Status1}{\color{white} \#24,columnize;}
				\colorbox{Status1}{\color{white} \#43,nes;}
				\colorbox{Status2}{\#804,xuperchain;}
				\colorbox{Status2}{\#255,qor;} 
				\colorbox{Status2}{\#780,ttn;}
				\colorbox{Status2}{\#6,ring;}
				\colorbox{Status2}{\#279,goczmq;}
				\colorbox{Status2}{\#334,bblfshd;}
				\colorbox{Status2}{\#333,sealos;}
				\colorbox{Status2}{\#239,pongo2;}
				\colorbox{Status2}{\#42,ccli;}
				\colorbox{Status2}{\#644,rqlite;}
				\colorbox{Status2}{\#629,direnv;}
				\colorbox{Status2}{\#3754,sensu-go;}
				\colorbox{Status2}{\#581,gost;}
				\colorbox{Status2}{\#181,cloud-game;}
				\colorbox{Status2}{\#313,gedcom;}
				\colorbox{Status2}{\#475,logspout;}
				\colorbox{Status2}{\#103,sdk;}
				\colorbox{Status2}{\#26,healthcheck;}
				\colorbox{Status2}{\#335,gostatsd;}
				\colorbox{Status2}{\#15,go-web-app;}
				\colorbox{Status2}{\#394,goproxy;}
				\colorbox{Status2}{\#26,go-corona;}
				\colorbox{Status2}{\#22,license;}
				\colorbox{Status2}{\#23,dque;}
				\colorbox{Status3}{\#2274,gobgp;}
				\colorbox{Status3}{\#1147,go-iost;}
				\colorbox{Status3}{\#72,goffmpeg;}
				\colorbox{Status3}{\#1272,go-algorand;}
				\colorbox{Status3}{\#16381,tidb;}
				\colorbox{Status3}{\#25,hyperfox;}
				\colorbox{Status3}{\#9,cuid;}
				\colorbox{Status5}{\#195,vaulted;}
				\colorbox{Status5}{\#561,moira;}
				\colorbox{Status5}{\#990,tidb;}
				\colorbox{Status5}{\#1747,vpp;}
				\colorbox{Status5}{\#95,hashi;}
				\colorbox{Status5}{\#43,jsonrpc;}
				\colorbox{Status5}{\#32,jsonrpc;}
				\colorbox{Status5}{\#333,goim;}
				\colorbox{Status5}{\#4,chive;}
				\colorbox{Status5}{\#90,go-rest-api;}
				\colorbox{Status5}{\#214,manba;}
				\colorbox{Status5}{\#4,openssl;}
				\colorbox{Status5}{\#59,go-arty;}
				\colorbox{Status5}{\#22,ynab.go;}
				\colorbox{Status5}{\#21,libgrin;}
				\colorbox{Status5}{\#727,bettercap;}
				\colorbox{Status5}{\#4,skl-go;}
				\colorbox{Status5}{\#293,sso;}
				\colorbox{Status5}{\#222,linx-server;}
				\colorbox{Status5}{\#306,k8s-adapter;}
				\colorbox{Status5}{\#212,go-nebulas;}
				\colorbox{Status5}{\#77,terminal;}
				\colorbox{Status5}{\#43,uiprogress;}
				\colorbox{Status5}{\#45,roger;}
				\colorbox{Status5}{\#37,gann;}
				\colorbox{Status5}{\#7,recaptcha;}
				\colorbox{Status5}{\#27,gnark;}
				\colorbox{Status5}{\#13,kratos-demo;}
				\colorbox{Status5}{\#1,metrics;}
				\colorbox{Status5}{\#16,gotypist;}
				\colorbox{Status5}{\#1,Goid;}
				\colorbox{Status5}{\#25,echo-session;}
				\\  \bottomrule
				\multicolumn{2}{l}{\begin{large} \colorbox{Status1}{\color{white} Status 1}: Issues fixed using our suggestions; \colorbox{Status2}{Status 2}: Issues under fixing using our suggestions; \colorbox{Status3}{Status 3}: Issues confirmed, but fixing not decided;
						\colorbox{Status4}{Status 4}: Issues fixed using other suggestions;\end{large}}\\
				\multicolumn{2}{l}{\begin{large}
						\colorbox{Status5}{Status 5}: Issues pending; Issue ID{\color{red}$\spadesuit$}: Migration to {\mycode Go Modules} conducted (desired); 
						Due to page limit, the detailed information of reported issues is provided on our homepage (\textbf{http://www.hero-go.com/})\end{large}.
				}\\[-0.06cm] 
			\end{tabular}
		}
	\end{center}
	\vspace{-5mm}
\end{table*}

In total, \textsc{Hero} reported 2,422 new issues after analyzing the 19,000 Golang projects.
Although the key information of root causes and fixing suggestions can be automatically generated by \textsc{Hero}, reporting these issues to developers involves substantial manual work, such as communicating with developers, helping them submit PRs, etc.
As such, we only managed to report 280 issues for the top 1001\textendash2000 popular projects (top 1\textendash1000 already used for RQs 2\textendash4) in the projects' issue trackers.
Table~3 summarizes the status of our reported issues.
Encouragingly, 181 issues (64.6\%) were quickly confirmed by the developers, and 160 confirmed issues (88.4\%) were later fixed or are under fixing. For all but two fixed issues, developers adopted our suggested fixes. The other issues are still pending (likely due to the inactive maintenance of the projects). We discuss the feedback from the developers below.


\textbf{\emph{Feedback on issue detection.}} 
While different types of DM issues had different confirmation rates (52.0\%\textendash74.4\%), most confirmed issues received positive feedback from developers. We give some examples below.
In issue \#2922 (\emph{Type A}) of {\mycode kiali}~\cite{Issue2922}, a developer mentioned 
``\emph{I have found the same issue as you describe via the commit {\mycode c453e89}~\cite{c453e89}. I just stuck in an older version of this library}''.
In issue \#256 (\emph{Type B.1}) of {\mycode flamingo-commerce}~\cite{Issue256}, developers were previously unaware of the risk and commented ``\emph{I guess the inconsistency of library version was imported by accident. We will create a PR to remove the occurrence}''.
In issue \#114 (\emph{Type B.2}) of {\mycode tomato}~\cite{Issue114}, a developer commented ``\emph{Nice catch! I think it is nice to clean up our vendor directory, since library {\mycode bitly/go-nsq} repository is not existed anymore}.''
We also reported issue \#16381 (\emph{Type C})~\cite{Issue16381} to project {\mycode tidb}~\cite{tidb} that violated SIV rules and the issue could affect 341 downstream projects!
Our report struck a chord with {\mycode tidb}'s downstream projects and was linked to seven real issues that indeed caused build failures (e.g., issue \#187~\cite{Issue187} of {\mycode parser}).

\textbf{\emph{Feedback on fixing suggestions.}}
To ease discussion, we divide the 160 DM issues that have been fixed or are under fixing into three categories: (1) 143 taking our highlighted preferred solutions (with minimal impacts to other projects), (2) 15 taking one of our suggestions (impacting some projects), and (3) the remaining two not taking our suggestions. 

As an example for category (1), issue \#3754~\cite{Issue3754} was induced by project {\mycode sensu-go}'s~\cite{sensu-go} SIV rule violations.
\textsc{Hero} warned the potential build errors for {\mycode sensu-go}'s 89 downstream projects.
This was confirmed by developers' comments ``\emph{We are aware of this issue, but the way you have summarized it, including the paths forward and impact analyses, is very valuable.}''
However, the developers could not follow SIV rules immediately due to some internal restrictions.
To minimize the impacts to these downstream projects, they tagged a ``\emph{technical-debt}'' to our report, and extracted part of the project code into a new module that follows SIV rules for use by downstream projects. This code refactoring process was laborious.
For category~(2), the developers did not take our highlighted preferred fixing solutions. With the information of impacted downstream projects reported by \textsc{Hero}, some developers chose to add notes in their projects' documentations to suggest the concerned downstream projects work around potential DM issues by using \emph{replace directives} (\emph{Solution 5}) or hash commit ID (\emph{Solution 8}) (e.g., issues \#16381 of {\mycode tidb}~\cite{Issue16381}).
For category (3), developers of only two reported issues (\#3970 of {\mycode sensu-go}~\cite{Issue3970} and \#770 of {\mycode libvirt}~\cite{Issue770}) did not take our fixing suggestions. Not wanting to be involved into trouble, they used other similar libraries for substitution.

The above feedback indicates that \textsc{Hero} is useful in detecting and predicting DM issues for Golang projects, as well as suggesting proper fixes with impact analysis.
Developers also showed interest in the \textsc{Hero} tool.
For example, one developer commented ``\emph{I found that you sent many contributions on GitHub for this kind of subjects on many repositories. How do you detect the problems with {\mycode Go Modules}? Do you plan to share a tool or something to manage {\mycode Go Modules} issues?}'' ({\mycode ovh/cds}'s~\cite{ovh/cds} issue \#5366~\cite{Issue5366}). Another commented ``\emph{It is a good bot!}'' ({\mycode TheThingsNetwork}'s issue \#780~\cite{Issue780}).
Encouraged by such comments, we are planning to release our tool for public use to help build a healthy Golang ecosystem.

\section{Discussions}
\label{sec:Discussions}

\subsection{Threats to Validity}
One possible threat is the representativeness of the studied Golang projects and DM issues. 
To reduce the threat, we selected top 20,000 projects on GitHub for migration status analysis (RQ1), and randomly chose 500 from the top 1,000 projects to investigate DM issues' characteristics (RQs~2\textendash3).
These projects are popular, large-sized, and well-maintained. We believe that they are proper subjects for our study.


Another possible threat is the generality of the issues that \textsc{Hero} detects since the issue types were observed by studying only 500 Golang projects.
To mitigate the threat, we used a different set of DM issues to evaluate \textsc{Hero} (RQ4) and found that \textsc{Hero} can detect 98.5\% of these issues, which suggests that our findings on issue characteristics are generalizable.
Besides, \textsc{Hero} also detected a large number of real DM issues after analyzing 19,000 Golang projects. This further suggests the generality of the findings in this paper.

In addition, our study involves manual work (e.g., identifying and analyzing issue reports).
To reduce the threat of human mistakes, three co-authors have cross-validated all results for consistency. 

\subsection{\textsc{Hero}'s Generalizability Beyond the Golang Ecosystem}

Two aspects of our methodology are generalizable to the DM issues induced by incompatible library-referencing modes at other ecosystems:

\begin{itemize}[leftmargin=*, topsep=0pt]
	\item The scenarios of issue types and their causes: (1) projects in the legacy library-referencing mode depend on projects in the new library-referencing mode, (2) projects in the new mode depend on those in the legacy mode, and (3) projects in the new mode depend on others also in the new  mode, can be generalized to analyze similar situations.
	
	\item The formulation of issue fixing patterns. The methodology to construct the dependency model by collecting information about its upstream and downstream projects can be adapted to other ecosystems. With the aid of such a dependency model, fixing suggestions can be structurally formulated based on applicable solutions and their potential impacts.
	The generalization of our methodology needs to consider the unique characteristics of the studied programming languages, since our work focuses only on the Golang ecosystem (one of the most influential and fastest growing open-source ecosystems). 
\end{itemize}



\section{Related Work}
\label{sec:Related work}

\textbf{\emph{Software dependency management.}}
Software dependency management has inherent complexities~\cite{Chang2020, dependenciesMSR19, ghorbani2019detection, patra2018conflictjs, wang2018dependency, wang2019could, wang2watchman, huang2020interactive, dig2006apis, henkel2005catchup, bavota2015apache, cox2015measuring, decan2018evolution, derr2017keep, kula2018developers, wang2020empirical, mccamant2003predicting, foo2018efficient, raemaekers2017semantic, raemaekers2012measuring, ruiz2016analyzing, kabinna2016logging, de2018library, kula2017exploratory, macho2018automatically, bezemer2017empirical, mostafa2017experience}.
Blincoe et al.~\cite{dependenciesMSR19} studied over 70 million dependencies to find out how developers declared dependencies across 17 package managers.
Their results guided research into better
practices for dependency management.
Abate et al.~\cite{ABATE20122228} reviewed state-of-the-art dependency managers and their ability to keep up with evolution at the current growth rate of popular component-based platforms, and conclude that their dependency solving abilities are not up to the task.
Some studies~\cite{bavota2015apache, cox2015measuring, decan2018evolution, derr2017keep, kula2018developers, wang2020empirical, mccamant2003predicting, foo2018efficient, raemaekers2017semantic, raemaekers2012measuring, ruiz2016analyzing, kula2017exploratory} focused on upgrading dependency versions, and some~\cite{dig2006apis, henkel2005catchup, kabinna2016logging, de2018library, mostafa2017experience} investigated how to migrate client code
to adapt to changing dependencies.
Researchers~\cite{patra2018conflictjs, wang2018dependency, wang2019could, wang2watchman, huang2020interactive} also proposed a series of techniques to detect, test and monitor dependency conflict issues (e.g., misusing versions) for JavaScript, Java, and Python projects.
Different from such conflict issues, our studied DM issues are due to incompatible library-referencing modes and their broad impacts on related projects in the Golang ecosystem.
Garcia et al.'s work~\cite{ghorbani2019detection} is closely related to our \textsc{Hero}, in which eight inconsistent modular dependencies were formally defined for Java-9 applications on the Java Platform Module System (JPMS).
They proposed a technique \textsc{Darcy} to detect and repair such inconsistencies but their targeted issues are architecture-implementation mapping ones, which are different from our focus.

\textbf{\emph{Health of software ecosystems.}}
Literatures on evolving software ecosystems cover {\mycode Maven}~\cite{soto2019emergence, benelallam2019maven, mitropoulos2014bug}, {\mycode Apache}~\cite{bavota2015apache, hernandez2015identifying}, {\mycode Eclipse}~\cite{businge2012survival}, {\mycode Ruby}~\cite{syeed2014socio, kabbedijk2011steering, kikas2017structure}, {\mycode PyPI}~\cite{wang2watchman}, {\mycode GNOME}~\cite{jergensen2011onion}, and {\mycode Npm}~\cite{trockman2018adding, zimmermann2019small, cogo2019empirical, staicu2020extracting, kikas2017structure, robles2018empirical, lertwittayatrai2017extracting, abdellatif2020simplifying}.
Many concerned techniques focus on three aspects: ecosystem modeling and analysis~\cite{blincoe2019reference, zimmermann2019small, cogo2019empirical, benelallam2019maven, hernandez2015identifying, kikas2017structure, lertwittayatrai2017extracting, abdellatif2020simplifying}, socio-technical theories within ecosystems~\cite{blincoe2019reference, trockman2018adding}, and diagnosis and monitoring for ecosystem's evolution~\cite{wang2watchman, jansen2014measuring, soto2019emergence}.
For example, Blincoe et al.~\cite{blincoe2019reference} proposed coupling references to model technical dependencies between projects, and explored characteristics of open-source or commercial software ecosystems.
Zimmermann et al.~\cite{zimmermann2019small} modeled dependencies for the {\mycode Npm} ecosystem, and analyzed potential risks for packages that could be attacked.
To the best of our knowledge, our work is the first attempt to study the health of Golang ecosystem from the perspective of DM issues.

\section{Conclusions and Future Work}
\label{sec:Conclusion}
In this paper, we studied DM issues in Golang projects, which are prevalent and have caused confusions and troubles to many Golang developers.
In particular, we investigated the characteristics of DM issues, analyzed their root causes, and identified common fixing solutions.
We refined our findings into detecting algorithms with customizable fixing templates.
The evaluation confirmed the effectiveness of our efforts as
a tool implementation \textsc{Hero} in detecting and diagnosing
DM issues. Leveraging fixing templates and rich diagnostic
information, we plan to study DM patch generation in future.


\section*{Acknowledgment}
The authors express thanks to the anonymous reviewers for their constructive comments. 
Part of the work was conducted during the first author's internship at HKUST in 2018. 
The work is supported by the National Natural Science Foundation of China (Grant Nos. 61932021, 61902056, 61802164, 61977014), Shenyang Young and Middle-aged Talent Support Program (Grant No. ZX20200272), the Fundamental Research Funds for the Central Universities (Grant No. N2017011), the Hong Kong RGC/GRF grant 16207120, MSRA grant, US NSF (Grant No. CCF-1845446) and Guangdong Provincial Key Laboratory (Grant No. 2020B121201001).

\balance
\bibliographystyle{IEEEtran}
\bibliography{bibliography}


\end{document}